\documentclass[%
 aps,
 jmp,%
 amsmath,amssymb,
 reprint,%
]{revtex4-1}
\usepackage{graphicx}
\usepackage{subfigure}
\usepackage{mathrsfs}
\usepackage[bookmarks=true,colorlinks=true,citecolor=blue,linkcolor=red,urlcolor=magenta]{hyperref}
\usepackage{multirow}
\usepackage{dcolumn}
\usepackage{bm}

\graphicspath{{figures_part2/}}

\begin{document}

\preprint{APS/123-QED}

\title[]{Mechanisms of proton-proton inelastic cross-section growth in multi-peripheral model within the framework of perturbation theory. Part 2}

\author{I.V. Sharf}
\affiliation{ Odessa National Polytechnic University, Shevchenko av. 1, Odessa, 65044, Ukraine.}%

\author{G.O. Sokhrannyi}%
\affiliation{ Odessa National Polytechnic University, Shevchenko av. 1, Odessa, 65044, Ukraine.}%

\author{A.V. Tykhonov}%
\affiliation{ Odessa National Polytechnic University, Shevchenko av. 1, Odessa, 65044, Ukraine.}%
\affiliation{ Department of Experimental Particle Physics, Jozef Stefan Institute,
     Jamova 39, SI-1000 Ljubljana, Slovenia.}%

\author{K.V. Yatkin}%
\affiliation{ Odessa National Polytechnic University, Shevchenko av. 1, Odessa, 65044, Ukraine.}%

\author{N.A. Podolyan}%
\affiliation{ Odessa National Polytechnic University, Shevchenko av. 1, Odessa, 65044, Ukraine.}%

\author{M.A. Deliyergiyev}%
\affiliation{ Odessa National Polytechnic University, Shevchenko av. 1, Odessa, 65044, Ukraine.}%
\affiliation{ Department of Experimental Particle Physics, Jozef Stefan Institute,
     Jamova 39, SI-1000 Ljubljana, Slovenia.}%

\author{V.D. Rusov}%
 \email{siiis@te.net.ua}
\affiliation{ Odessa National Polytechnic University, Shevchenko av. 1, Odessa, 65044, Ukraine.}%
\affiliation{Department of Mathematics, Bielefeld University, 
      Universitatsstrasse 25, 33615 Bielefeld, Germany.}%


\date{\today}

\begin{abstract}
We demonstrate a new technique for calculating proton-proton inelastic cross-section, which allows one by application of the Laplace' method replace the integrand in the integral for the scattering amplitude in the vicinity of the maximum point by expression of Gaussian type. This in turn, allows one to overcome the computational difficulties for the calculation of the integrals expressing the cross section to sufficiently large numbers of particles. We have managed to overcome these problems in calculating the proton-proton inelastic cross-section for production ($n \le 8$) number of secondary particles in within the framework of ${\phi ^3}$ model. As the result the obtained dependence of inelastic cross-section and total scattering cross-section on the energy $\sqrt s $ are qualitative agrees with the experimental data. Such description of total cross-section behavior differs considerably from existing now description, where reggeons exchange with the intercept greater than unity is considered.
\end{abstract}

\keywords{ inelastic scattering cross-section, total scattering cross-section, Laplace method, virtuality, multi-peripheral model, Regge theory}
\maketitle

%

\section{Introduction}
\label{INTRO}

The problems of the inelastic scattering cross-sections calculation have been discussed in details in [\onlinecite{part1}]. As the result of approximations, which are usually made to overcome these difficulties [\onlinecite{springerlink:10.1007/BF02781901, Holliday_NuovoCimento, bfkl_1976, Collins:111502}], are obtained the integral over the areas of phase space, where different points correspond to different values of the energy-momentum, but at the same time they come to the equation with equal weights. Therefore, the energy-momentum conservation law does not consider reasonably. 

Besides, the virtualities, in the equation for the amplitude, reduced to values of the square of transversal components particles momentums [\onlinecite{Lipatov:2004}], meanwhile the rest components of virtualities are not insignificant and appear as quite essential [\onlinecite{part1}].

These approximations are based on the assumption that the main contribution to the integral makes the multi-Regge domain [\onlinecite{Lipatov:2008}]. This assumption is crucial for the modern approaches to the description of inelastic scattering processes [\onlinecite{KozlovNSU_2007}]. However, the obtained results in [\onlinecite{part1}] led to the conclusion that main contribution in the integral does not make the multi-Regge domain.

The aim of this paper is to propose an alternative method for calculating inelastic scattering cross-sections based on well-known the Laplace method for the multidimensional integral [\onlinecite{DeBruijn:225131}]. In order to apply this method it is required the element of integration has the point of maximum within integration domain. It has been shown [\onlinecite{part1}], that for the diagrams of ``comb" type with the accurate energy-momentum conservation law calculation the square of scattering amplitude module is really has that maximum. 

Analysis of the properties of this maximum led to the conclusion that there is the mechanism of cross-section growth. This mechanism has not been considered previously, due to the above approximations associated with the multi-Regge kinematics. Now we would like to show that this mechanism can be responsible for the experimentally observed behavior of cross sections dependence with energy $\sqrt s$. Actually, application of Laplace's method to the processes of production of large number of secondary particles faces the challenge of accounting the vast amount of interference contributions, which will be discussed in detail in Section.\ref{SECTION_2}. In fact, there are $n!$ of these contributions for the process with production of $n$ secondary particles. Therefore, in the present paper we were able to calculate all interference contributions to the production of $n\le8$ secondary particles. 

Typically, these contributions are underestimate, because according to the considering assumption, that particles on the ``comb"  strongly ordered in rapidity [\onlinecite{Holliday_NuovoCimento}] or strongly ordered in Sudakov`s parameters [\onlinecite{bfkl_1976}], they should be negligible.

However, later in this paper, we will show, that these contributions are significant and the contribution from the square modulus of just one ``comb" type diagram with initial particles arrangement, which are usually limited,  is only a small fraction of the sums of all interference contributions. Despite the fact, that the partial cross sections were calculated only for a small number of secondary particles, we were able to achieve a qualitative agreement with experimental results.

\section{On the need of consideration of diagrams with the different sequence of the attaching external lines to the ``comb"}
\label{SECTION_1}
An inelastic scattering cross-section, which is interesting for us, is described by the following equation:
\begin{eqnarray}
 && \mbox{\fontsize{14}{14}\selectfont  ${\sigma _n} = \frac{1}{4n!I} \int {\frac{{d{{\vec P}_3}}}{{2{P_{30}}{{(2\pi )}^3}}}} \frac{{d{{\vec P}_4}}}{{2{P_{40}}{{(2\pi )}^3}}}\prod\limits_{k = 1}^n {\frac{{d{{\vec p}_k}}}{{2{p_{0k}}{{(2\pi )}^3}}}   }$ } \nonumber\\
 && \mbox{\fontsize{11}{14}\selectfont $ \times \Phi {\delta ^{\left( 4 \right)}}\left (  {{P_3} + {P_4} + \sum\limits_{k = 1}^n {{p_k}}  - {P_1} - {P_2}} \right) $ }
\label{eq1}  
\end{eqnarray}
\begin{subequations}
\begin{eqnarray}
&&I = \sqrt {{{({P_1}{P_2})}^2} - {{({M_1}{M_2})}^2}}
\\
&&\Phi = {\left| {T(n,{p_1},{p_2},...,{p_n},{P_1},{P_2},{P_3},{P_4})} \right|^2}
\end{eqnarray}
\end{subequations}

The scattering amplitude in this equation will be considered within framework of the multi-peripheral model, i.e., for the diagrams of ``comb" type. However, here we will make the important remark.

\begin{figure}
\begin{center}
\subfigure[]{
  \includegraphics[scale=0.65]{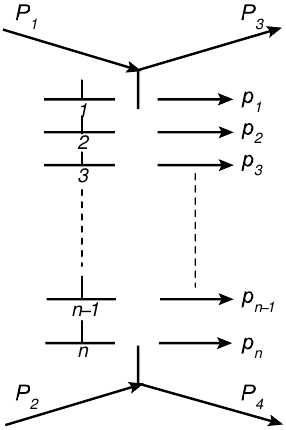}
  \label{fig:fig_part2_01a} 
}
\subfigure[]{
  \includegraphics[scale=0.65]{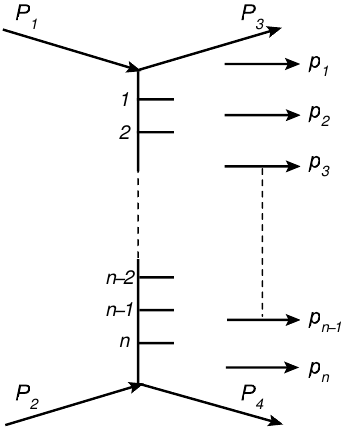}
  \label{fig:fig_part2_01b} 
}
\end{center}
\caption{Drawing the diagrams of the ``comb'' type. }
\label{fig:fig_part2_01}
\end{figure}

According to the Wick theorem, the scattering amplitude is the sum of diagrams with all possible orders of external lines attaching to the ``comb". In the terms of diagram technique it looks as follows. Plotting the multi-peripheral diagram of the scattering amplitude (as it is shown in Fig.\ref{fig:fig_part2_01} of [\onlinecite{part1}]) at first we have adequate number of vertices with three lines going out of it and $n$ lines corresponding to the secondary particles as it is shown in Fig.\ref{fig:fig_part2_01a}.

``Pairing" some lines Fig.\ref{fig:fig_part2_01a} in order to obtain the ``comb", we will get a situation shown in Fig.\ref{fig:fig_part2_01b}. The weighting coefficient appearing from this procedure, is included to the coupling constant. Finally we have to ``pair" the appropriate lines of particles in the final state with the remaining unpaired internal lines in the diagram of Fig.\ref{fig:fig_part2_01}.

If we marked by $i_1$ - the external line, paired with the first vertex; $i_2$ - the external line, paired with the second vertex and etc.; then $i_k$ is an external line, which is paired with $k$-th vertex, so every diagram will be characterized by sequence $i_1, i_2,\ldots,i_n$. And in this case the total amplitude is expressed by the sum of $n!$ terms, each of them corresponds to one of $n!$ possible index sequences and therefore the inelastic scattering cross-section can be written as 
\begin{eqnarray}
 && \mbox{\fontsize{14}{14}\selectfont  ${\sigma _n} = I'\int {\frac{{d{{\vec P}_3}}}{{2{P_{30}}{{(2\pi )}^3}}}} \frac{{d{{\vec P}_4}}}{{2{P_{40}}{{(2\pi )}^3}}}\prod\limits_{k = 1}^n {\frac{{d{{\vec p}_k}}}{{2{p_{0k}}{{(2\pi )}^3}}}   }$ } \nonumber\\
 && \mbox{\fontsize{11}{14}\selectfont $ \times \Phi'{\delta ^{\left( 4 \right)}}\left (  {{P_3} + {P_4} + \sum\limits_{k = 1}^n {{p_k}}  - {P_1} - {P_2}} \right) $ }
\label{eq2}  
\end{eqnarray}
\begin{subequations}
\begin{eqnarray}
&& I' = \frac{{{{\left( {{{\left( {2\pi } \right)}^4}} \right)}^2}{g^4}{\lambda ^{2n}}}}{{4n!\sqrt {{{({P_1}{P_2})}^2} - {{({M_1}{M_2})}^2}} }} \label{eq3a} \\
&& \mbox{\fontsize{8}{14}\selectfont $ \Phi' = {\left( {\sum\limits_{P\left( {{i_1},{i_2},...,{i_n}} \right)} {A\left( {n,{P_3},{P_4},{p_{{i_1}}},{p_{{i_2}}},...,{p_{{i_n}}},{P_2},{P_1}} \right)} } \right)^*}$ } \nonumber\\ 
&& \mbox{\fontsize{8}{14}\selectfont $ \times \left( {\sum\limits_{P\left( {{j_1},{j_2},...,{j_n}} \right)} {A\left( {n,{P_3},{P_4},{p_{{j_1}}},{p_{{j_2}}},...,{p_{{j_n}}},{P_2},{P_1}} \right)} } \right)$ }
\label{eq3b} 
\end{eqnarray}
\end{subequations}

Here, as well as in [\onlinecite{part1}], ${M_1}$ and ${M_2}$ are the masses of particles in initial state and we assume that ${M_1} = {M_2} = M$, where $M$ is proton mass. Moreover, ${P_1}$ and ${P_2}$ are four-momenta of initial protons; ${P_3}$ and ${P_4}$ are four-momenta of protons in the final state; ${p_k},k = 1,2,...n$ are four-momenta of secondary particles (pions of mass $m$). As the virtual particles we understand the quanta of real scalar field with pion mass $m$. A coupling constant in vertexes, in which the pion lines join with proton lines, is denote as $g$ and $\lambda $ is coupling constant in vertexes, where three pion lines meet. The function $A$ is defined by: 
\begin{eqnarray}
&& \mbox{\fontsize{12}{14}\selectfont $ A\left( {n,{P_3},{P_4},{p_1},{p_2},...,{p_n},{P_1},{P_2}} \right) =  $ } \nonumber\\  
&& \mbox{\fontsize{12}{14}\selectfont $  = \frac{1}{{{m^2} - {{\left( {{P_1} - {P_3}} \right)}^2} - i\varepsilon }} $ } \nonumber\\ 
&& \mbox{\fontsize{12}{14}\selectfont $  \times \frac{1}{{{m^2} - {{\left( {{P_1} - {P_3} - {p_1}} \right)}^2} - i\varepsilon }} $ } \nonumber\\ 
&& \mbox{\fontsize{12}{14}\selectfont $  \times \frac{1}{{{m^2} - {{\left( {{P_1} - {P_3} - {p_1} - {p_2}} \right)}^2} - i\varepsilon }} \cdots  \cdots  \cdots $ } \nonumber\\  
&& \mbox{\fontsize{12}{14}\selectfont $  \times \frac{1}{{{m^2} - {{\left( {{P_1} - {P_3} - {p_1} - {p_2} - ... - {p_{n - 1}}} \right)}^2} - i\varepsilon }} $ } \nonumber\\ 
&& \mbox{\fontsize{12}{14}\selectfont $  \times \frac{1}{{{m^2} - {{\left( {{P_1} - {P_3} - {p_1} - {p_2} - ... - {p_{n - 1}} - {p_n}} \right)}^2} - i\varepsilon }} $ }
\label{eq3c}
\end{eqnarray}
Moreover, as it was shown in [\onlinecite{part1}], the function $A$ is real and positive therefore sign of complex conjugation in Eq.\ref{eq2} can be dropped and we can rewrite this expression in the following form:
\begin{eqnarray}
 && \mbox{\fontsize{14}{14}\selectfont  $ {\sigma _n} = I'\sum\limits_{\scriptstyle P\left( {{i_1},{i_2}, \cdots ,{i_n}} \right) \hfill \atop 
  \scriptstyle P\left( {{j_1},{j_2}, \cdots ,{j_n}} \right) \hfill} {\int {\frac{{d{{\vec P}_3}}}{{2{P_{30}}{{\left( {2\pi } \right)}^3}}}\frac{{d{{\vec P}_4}}}{{2{P_{40}}{{\left( {2\pi } \right)}^3}}}} }   $ } \nonumber\\
 && \mbox{\fontsize{11}{14}\selectfont  $  \times \prod\limits_{k = 1}^n {\frac{{d{{\vec p}_k}}}{{2{p_{0k}}{{\left( {2\pi } \right)}^3}}}}  
\times {\delta ^{\left( 4 \right)}}\left( {{P_3} + {P_4} + \sum\limits_{k = 1}^n {{p_k}}  - {P_1} - {P_2}} \right) $ } \nonumber\\
 && \mbox{\fontsize{13}{14}\selectfont  $  \times A\left( {n,{P_3},{P_4},{p_{{i_1}}},{p_{{i_2}}}, \cdots ,{p_{{i_n}}},{P_2},{P_1}} \right) $ } \nonumber\\
 && \mbox{\fontsize{13}{14}\selectfont  $  \times A\left( {n,{P_3},{P_4},{p_{{j_1}}},{p_{{j_2}}}, \cdots ,{p_{{j_n}}},{P_2},{P_1}} \right) $ }
\label{eq5}
\end{eqnarray}
where $I'$ defined by Eq.\ref{eq3a}.

Notation $\sum\limits_{P\left( {{i_1},\;{i_2},...,\;{i_n}} \right)} {} $ means that we consider sum of terms corresponding to all possible permutations of indices ${i_1},{i_2},...,{i_n}$. Let us note that the integration variables in each of term of considered sum can be renaming, so that the indexes ${i_1},{i_2},...,{i_n}$ formed the original placing $1, 2,...,n$. At the same time the indexes ${j_1},{j_2},...,{j_n}$ will run through all possible permutations and summation must be carried over all these permutations. Taking into account this, we get instead of Eq.\ref{eq5}: 
\begin{eqnarray}
&&  {\sigma _n} = I''\int {\frac{{d{{\vec P}_3}}}{{2{P_{30}}{{\left( {2\pi } \right)}^3}}}\frac{{d{{\vec P}_4}}}{{2{P_{40}}{{\left( {2\pi } \right)}^3}}}\prod\limits_{k = 1}^n {\frac{{d{{\vec p}_k}}}{{2{p_{0k}}{{\left( {2\pi } \right)}^3}}}} }  \nonumber\\ 
 && \times {\delta ^{\left( 4 \right)}}\left( {{P_3} + {P_4} + \sum\limits_{k = 1}^n {{p_k}}  - {P_1} - {P_2}} \right)\nonumber\\ 
 &&  \times \Phi\left( {n,{P_3},{P_4},{p_1},{p_2}, \cdots ,{p_n},{P_2},{P_1}} \right) 
\label{eq5a}  
\end{eqnarray}
where 
\begin{eqnarray}
&& I'' = \frac{{{{\left( {{{\left( {2\pi } \right)}^4}} \right)}^2}{g^4}{\lambda ^{2n}}}}{{4\sqrt {s/4 - {{({M_1}{M_2})}^2}} \sqrt s }}
\label{eq6a} \\
 && \mbox{\fontsize{12}{14}\selectfont $ \Phi\left( {n,{P_3},{P_4},{p_1},{p_2}, \cdots ,{p_n},{P_2},{P_1}} \right)=$}\nonumber\\ 
&& \mbox{\fontsize{12}{14}\selectfont $ = A\left( {n,{P_3},{P_4},{p_1},{p_2},...,{p_n},{P_2},{P_1}} \right) $}\nonumber\\ 
&& \mbox{\fontsize{9}{14}\selectfont $ \times \sum\limits_{P\left( {{j_1},{j_2},...,{j_n}} \right)} {A\left( {n,{P_3},{P_4},{p_{{j_1}}},{p_{{j_2}}},...,{p_{{j_n}}},{P_2},{P_1}} \right)}$ } 
\label{eq6} 
\end{eqnarray}

Now we can use the fact that the amplitudes $A$ in Eq.\ref{eq6} have the points of constrained maximum [\onlinecite{part1}].

\section{Computation of multi-peripheral diagram contributions to inelastic scattering cross-section by Laplace’s method}
\label{SECTION_2}

We consider Eq.\ref{eq5} in the c.m.s. framework. Expanding the three-dimensional particle momenta into longitudinal and
transverse components with respect to the collision axis gives:
\begin{eqnarray}
&& \mbox{\fontsize{10}{14}\selectfont $ {\sigma _n} = I''\int {\frac{{d{{\vec P}_{3 \bot }}d{P_{3\left\| {} \right.}}}}{{2{{(2\pi )}^3}\sqrt {{M^2} + P_{3\left\| {} \right.}^2 + \vec P_{3 \bot }^2} }}} \frac{{d{{\vec P}_{4 \bot }}d{P_{4\left\| {} \right.}}}}{{2{{(2\pi )}^3}\sqrt {{M^2} + P_{4\left\| {} \right.}^2 + \vec P_{4 \bot }^2} }}$ } \nonumber\\ 
&& \mbox{\fontsize{10}{14}\selectfont $ \times \prod\limits_{k = 1}^n {\frac{{d{{\vec p}_{k \bot }}d{p_{k\left\| {} \right.}}}}{{2{{(2\pi )}^3}\sqrt {{m^2} + p_{k\left\| {} \right.}^2 + \vec p_{k \bot }^2} }}}$ } \nonumber\\
&& \mbox{\fontsize{8}{14}\selectfont $ \times \Phi\left( {n,{p_{1\parallel }},{{\vec p}_{1 \bot }}, \ldots ,{p_{n\parallel }},{{\vec p}_{n \bot }},{P_{1\parallel }},{P_{2\parallel }},{P_{3\parallel }},{{\vec P}_{3 \bot }},{P_{4\parallel }},{{\vec P}_{4 \bot }}} \right)$ } \nonumber\\
&& \mbox{\fontsize{10}{14}\selectfont $ \times \delta \left( {{T_3} + {T_4} + \sum\limits_{k = 1}^n {\sqrt {1 + p_{3\left\| {} \right.}^2 + \vec p_{3 \bot }^2} }  - \sqrt s } \right)$ } \nonumber\\ 
&& \mbox{\fontsize{9}{14}\selectfont $ \times \delta \left( {\sum\limits_{k = 1}^n {{p_{k\left\| {} \right.}}}  + {P_{3\left\| {} \right.}} + {P_{4\left\| {} \right.}}} \right)\delta \left( {\sum\limits_{k = 1}^n {{p_{k \bot x}}}  + {P_{3 \bot x}} + {P_{4 \bot x}}} \right)$ } \nonumber\\ 
&& \mbox{\fontsize{10}{14}\selectfont $ \times \delta \left( {\sum\limits_{k = 1}^n {{p_{k \bot y}}}  + {P_{3 \bot y}} + {P_{4 \bot y}}} \right)$ }
\label{eq7} 
\end{eqnarray}
where $I''$ defined by Eq.\ref{eq6a} and 
\begin{subequations}
\begin{eqnarray}
&& {T_3} = \sqrt {{M^2} + P_{3\parallel }^2 + \vec P_{3 \bot }^2}  \\ 
&& {T_4} = \sqrt {{M^2} + P_{4\parallel }^2 + \vec P_{4 \bot }^2}  
\end{eqnarray}
\end{subequations}
\begin{figure}
\begin{center}
\subfigure[]{
  \includegraphics[scale=0.23]{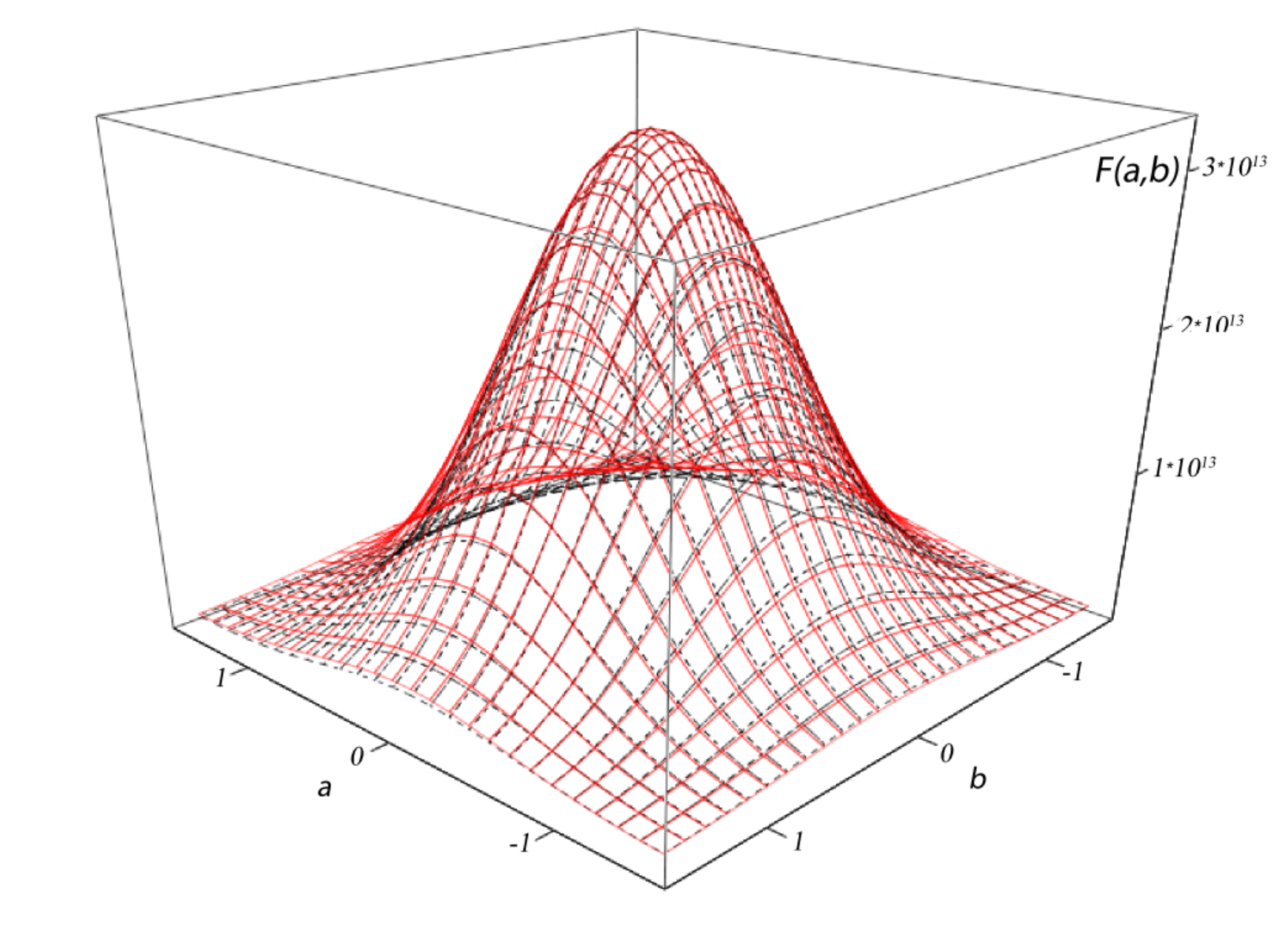}
  \label{fig:fig_part2_02a} 
}
\subfigure[]{
  \includegraphics[scale=0.23]{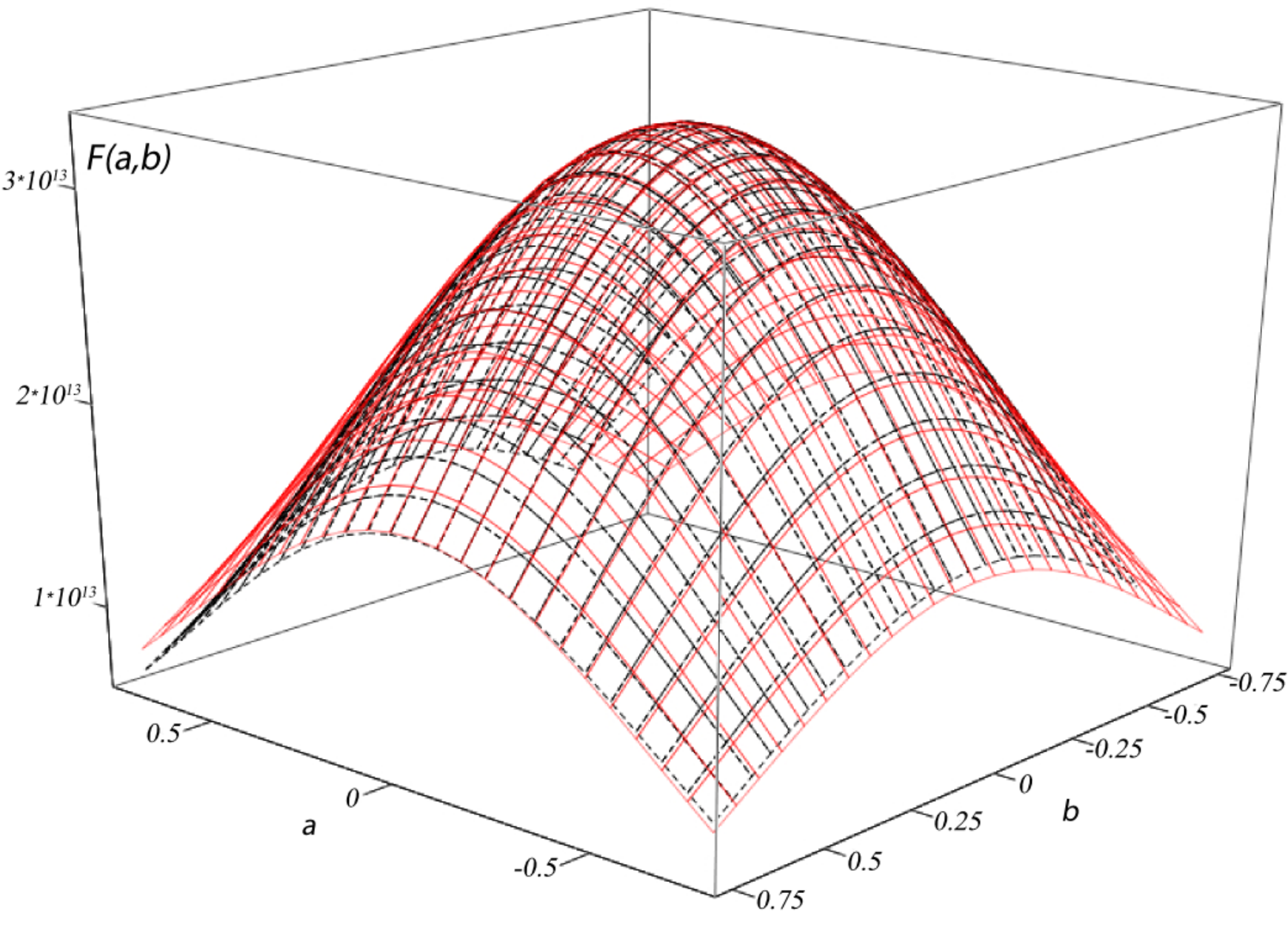}
  \label{fig:fig_part2_02b} 
}
\end{center}
\caption{
Functions $F_{1,7}^{n=10}(a,b)$ (dashed line) and $F_{1,7}^{(g), n=10}(a,b)$ (solid line) at energy $\sqrt s=5$ GeV for $n=10$. The general image \ref{fig:fig_part2_02a}, and the zoomed image \ref{fig:fig_part2_02b} is given in the vicinity of the point of maximum. Clear, that in region, which makes the most significant contribution to the integral, the scattering amplitude does not differ from its Gaussian approximation Eq.\ref{eq14}. This demonstrates the admissibility of applying the Laplace method.}
\label{fig:part2_fig02}
\end{figure}
\begin{figure}
\begin{center}
\subfigure[]{
  \includegraphics[scale=0.23]{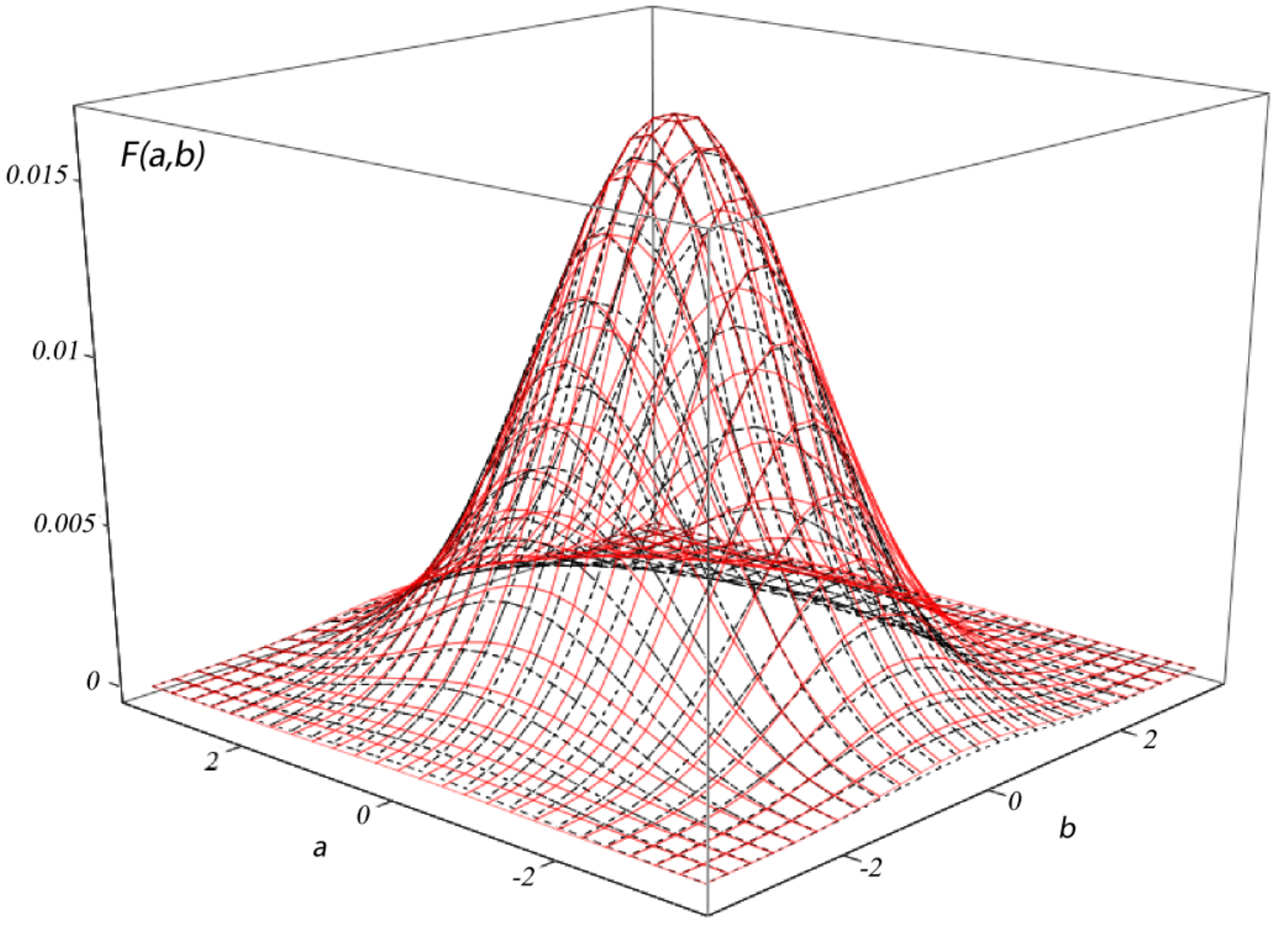}
  \label{fig:fig_part2_03a} 
}
\subfigure[]{
  \includegraphics[scale=0.23]{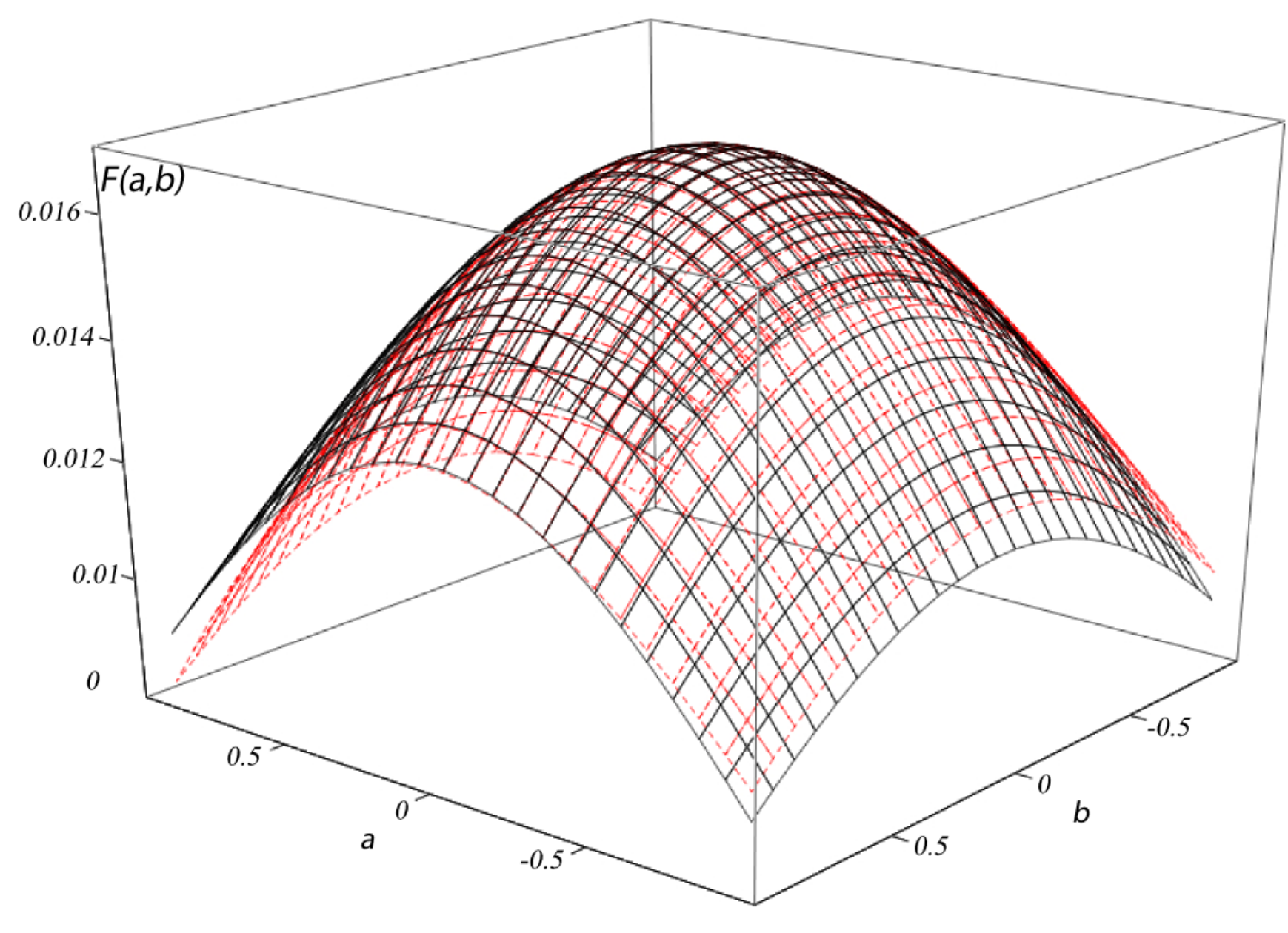}
  \label{fig:fig_part2_03b} 
}
\end{center}
\caption{
Functions $F_{1,7}^{n=10}(a,b)$ (dashed line) and $F_{1,7}^{(g), n=10}(a,b)$ (solid line) at energy $\sqrt s=100$ GeV for $n=10$. The general image \ref{fig:fig_part2_02a}, and the zoomed image \ref{fig:fig_part2_02b} is given in the vicinity of the point of maximum. Clear, that in region, which makes the most significant contribution to the integral, the scattering amplitude does not differ from its Gaussian approximation Eq.\ref{eq14}. This demonstrates the admissibility of applying the Laplace method.}
\label{fig:part2_fig03}
\end{figure}
The three $\delta$-functions in Eq.\ref{eq7}, whose arguments are linear with respect to integration variables, can be dropped out by integrating over $P_{4\parallel}$, $P_{4\perp x}$, $P_{4\perp y}$. The remaining  $\delta$-functions, which expresses the energy conservation law, can be computed by replacing $P_{3\parallel}$ with a new integration variable:
\begin{eqnarray}
&& {E_p} = \sqrt {{M^2} + P_{3\left\| {} \right.}^2 + \vec P_{3 \bot }^2}  \nonumber\\ 
&& \mbox{\fontsize{9}{10}\selectfont $ + \sqrt {{M^2} + {{\left( {\sum\limits_{k = 1}^n {{p_{k\left\| {} \right.}}}  + {P_{3\left\| {} \right.}}} \right)}^2} + {{\left( {\sum\limits_{k = 1}^n {{{\vec p}_{k \bot }}}  + {{\vec P}_{3 \bot }}} \right)}^2}}$ } 
\label{eq9}  
\end{eqnarray}
To make the following replacement we must express $P_{3\parallel}$ through $E_p$. The corresponding relation will coincide with Eq.8 in [\onlinecite{part1}] with the positive sign in front of the square root. Moreover, let’s introduce the rapidities instead of longitudinal momenta:
\begin{subequations}
\begin{eqnarray}
&& {p_{k\parallel }} = {m_ \bot }\left( {{{\vec p}_{k \bot }}} \right)sh\left( {{y_k}} \right) \\ 
&& {m_ \bot }\left( {{{\vec p}_{k \bot }}} \right) = \sqrt {m + \vec p_{k \bot }^2}  
\end{eqnarray}
\end{subequations}

After these transformations we have
\begin{eqnarray}
&& \mbox{\fontsize{14}{14}\selectfont $ {\sigma _n} = \frac{{{{\left( {2\pi } \right)}^2}{g^4}{\lambda ^{2n}}}}{{4\sqrt {s/4 - {M^2}} \sqrt s }}\int {\frac{{d{{\vec P}_{3 \bot }}}}{{2\sqrt {{M^2} + P_{3\parallel }^2 + \vec P_{3 \bot }^2} }}}$ }  \nonumber\\ 
&&  \times {\left. {\prod\limits_{k = 1}^n {\frac{{d{{\vec p}_{k \bot }}d{y_{k\parallel }}}}{{2{{(2\pi )}^3}}}\frac{{\partial {P_{3\parallel }}}}{{\partial {E_p}}}} } \right|_{{E_p} = \sqrt s  - \sum\limits_{k = 1}^n {{m_{ \bot k}}\left( {{{\vec p}_{ \bot k}}} \right)} ch\left( {{y_k}} \right)}} \nonumber\\ 
&& \mbox{\fontsize{13}{12}\selectfont $ \times \frac{{\Phi'}}{{2\sqrt {{M^2} + {{\left( {\sum\limits_{k = 1}^n {{p_{k\parallel }}}  + {P_{3\parallel }}} \right)}^2} + {{\left( {\sum\limits_{k = 1}^n {{{\vec p}_{k \bot }}}  + {{\vec P}_{3 \bot }}} \right)}^2}} }}$ } 
\label{eq11} 
\end{eqnarray}
where
\begin{eqnarray}
&& \mbox{\fontsize{12}{10}\selectfont $\Phi'=\Phi\left( {n,{y_1},{{\vec p}_{1 \bot }}, \ldots ,{y_n},{{\vec p}_{n \bot }}} \right., $}\nonumber\\ 
&& \mbox{\fontsize{10}{10}\selectfont $ \left. {{P_{1\parallel }},{P_{2\parallel }},{P_{3\parallel }},{{\vec P}_{3 \bot }},{P'_{4\parallel }},{{\vec P'}_{4 \bot }}} \right)$}
\label{eq11a} 
\end{eqnarray}
with
\begin{subequations}
\begin{eqnarray}
&& { P'_{4\parallel }} =  - \left( {\sum\limits_{k = 1}^n {{m_ \bot }\left( {{{\vec p}_{k \bot }}} \right)} {\mathop{\rm sh}\nolimits} \left( {{y_k}} \right) + {P_{3\parallel }}} \right) \\
&& {{\vec P'}_{4 \bot }}=  - \left( {\sum\limits_{k = 1}^n {{{\vec p}_{k \bot }}}  + {{\vec P}_{3 \bot }}} \right)
\end{eqnarray}
\end{subequations}
\begin{figure*}
\begin{center}
\subfigure[]{
  \includegraphics[scale=0.27]{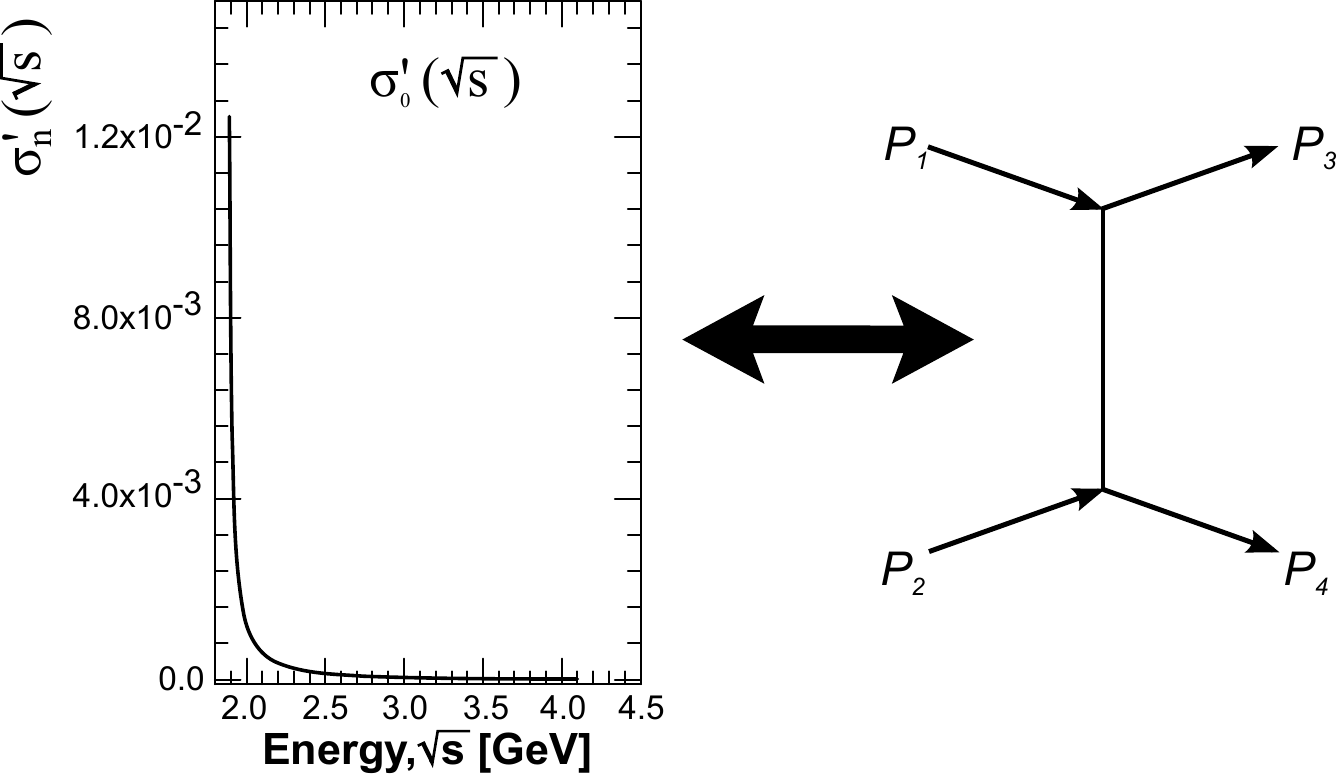} 
  \label{fig:part2_fig04a} 
}
\subfigure[]{
  \includegraphics[scale=0.27]{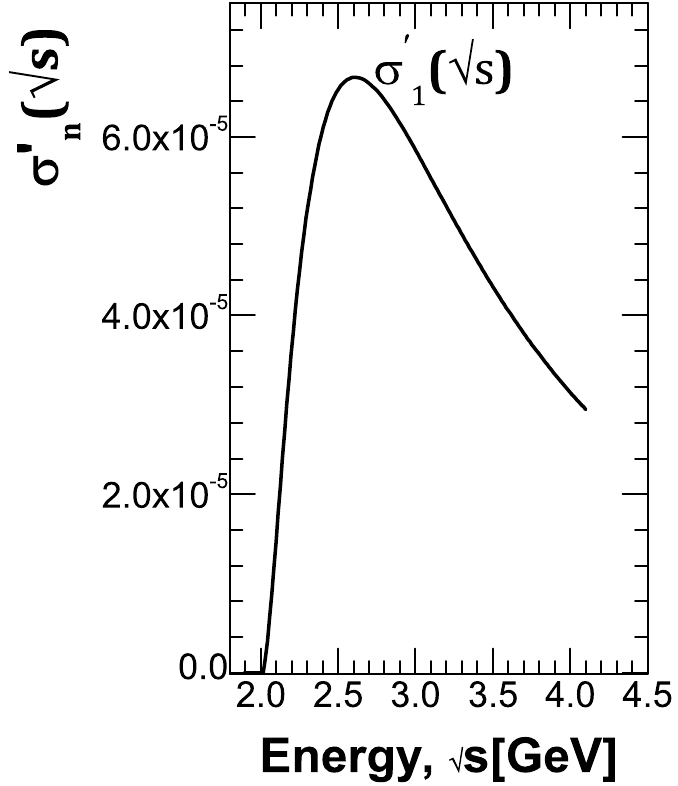} 
  \label{fig:part2_fig04b} 
}
\subfigure[]{
  \includegraphics[scale=0.27]{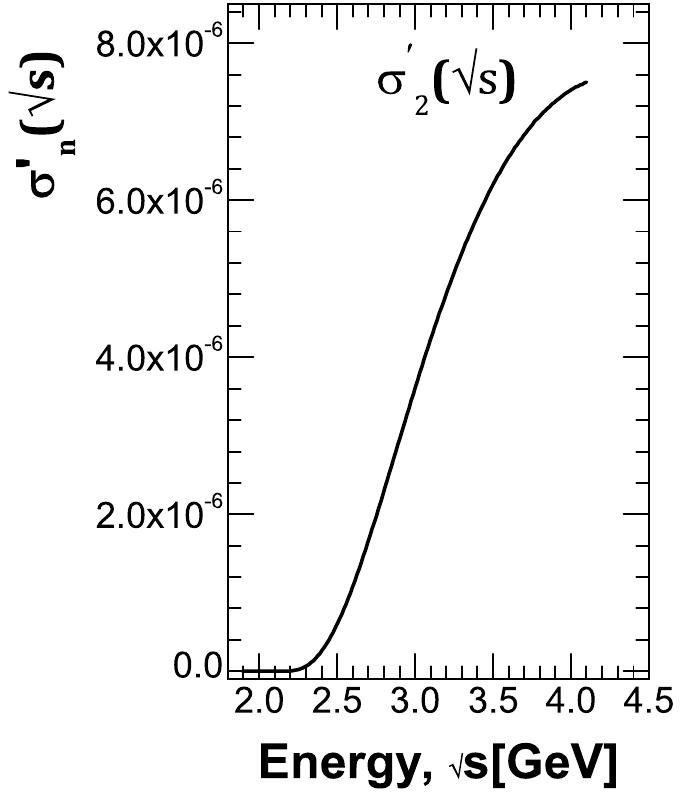} 
  \label{fig:part2_fig04c} 
}
\subfigure[]{
  \includegraphics[scale=0.27]{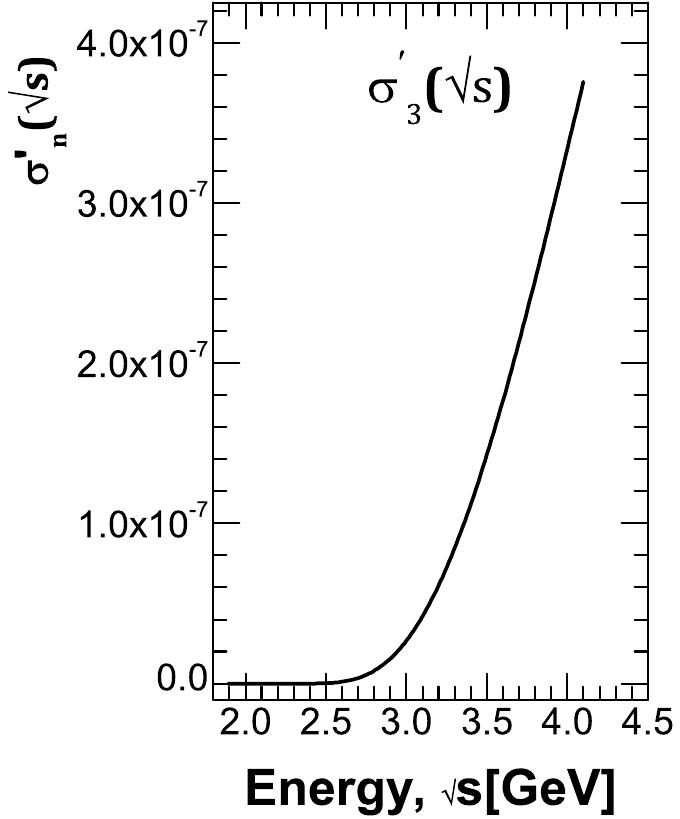} 
  \label{fig:part2_fig04d} 
}
\subfigure[]{
  \includegraphics[scale=0.27]{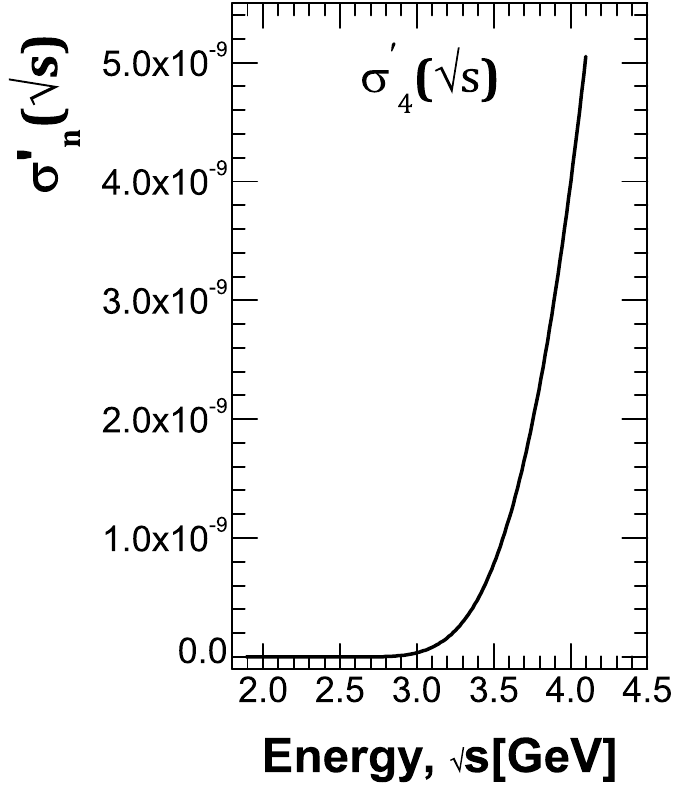} 
  \label{fig:part2_fig04e} 
}
\subfigure[]{
  \includegraphics[scale=0.27]{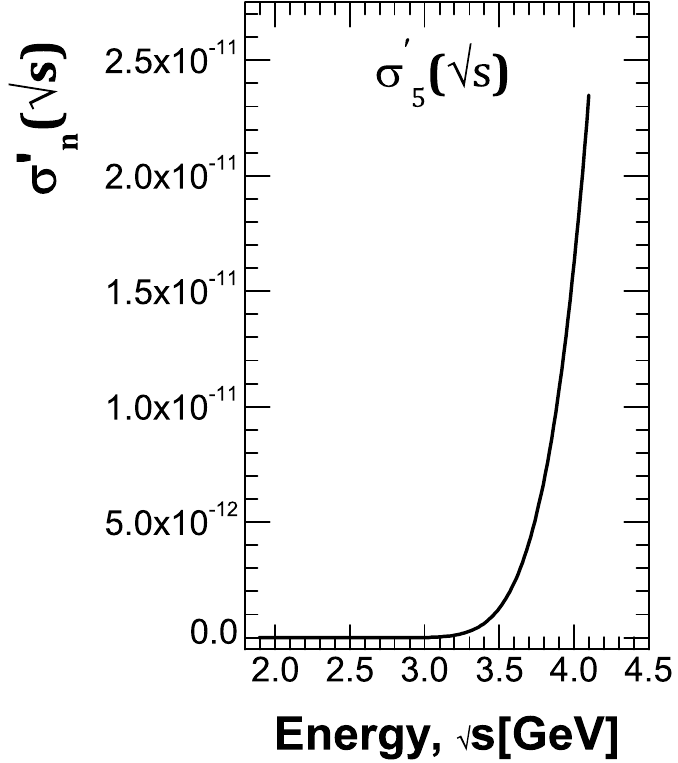} 
  \label{fig:part2_fig04f} 
}
\subfigure[]{
  \includegraphics[scale=0.27]{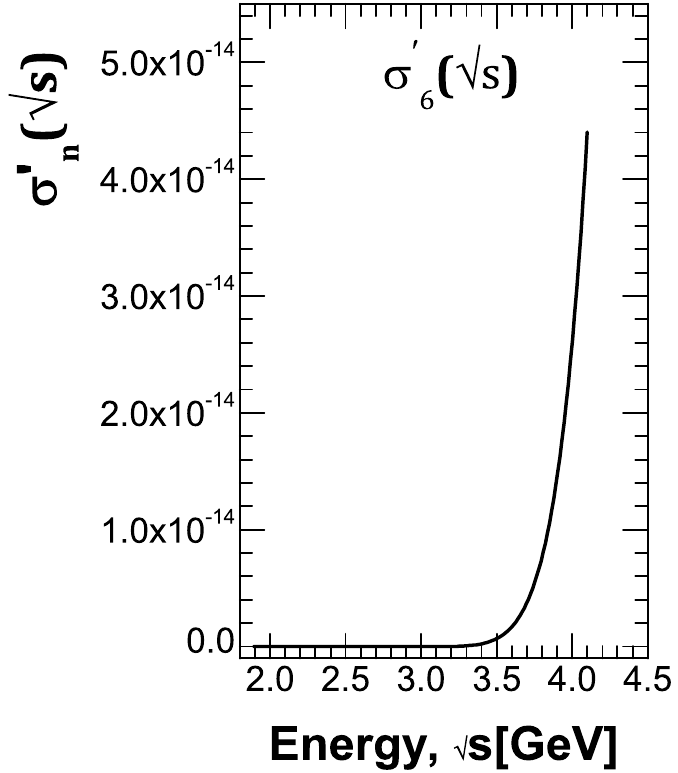} 
  \label{fig:part2_fig04g} 
}
\subfigure[]{
  \includegraphics[scale=0.27]{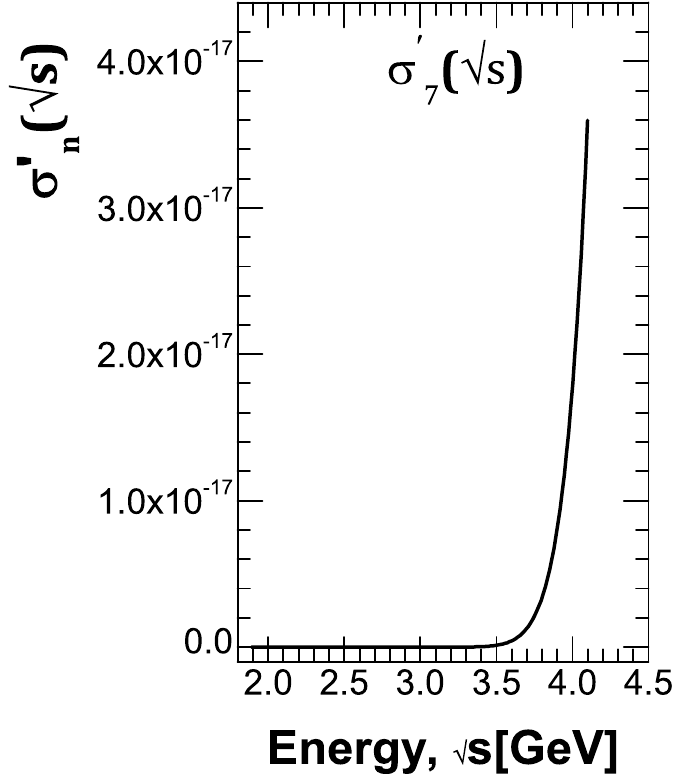} 
  \label{fig:part2_fig04h} 
}
\subfigure[]{
  \includegraphics[scale=0.27]{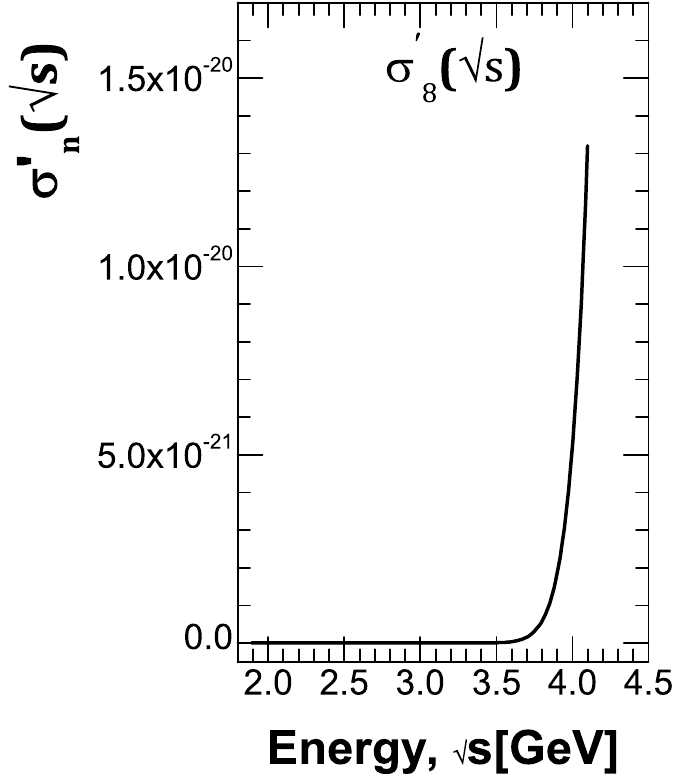} 
  \label{fig:part2_fig04i} 
}
\end{center}
\caption{
The values of $\sigma '_n(\sqrt s)$, $n=0,1,\ldots,8$ calculated in the range of threshold energies for $1, 2,\ldots,8$ particle productions. Via $\sigma '_0(\sqrt s)$ was denoted one of the contributions from the diagram (shown on the right) to inelastic scattering cross-section.}
\label{fig:fig_part2_04}
\end{figure*}
Note, that the magnitude ${P_{3\parallel }}$ is expressed in terms of the other integration variables via Eq.8 in [\onlinecite{part1}]. 

Next we turn to the dimensionless integration variables and make the following replacements: $\vec{p}_{k\perp} \to \frac{\vec{p}_{k\perp}}{m}$, $\vec{P}_{3\perp} \to \frac{\vec{P}_{3\perp}}{m}$. We will refer to the new dimensionless integration variables with the same notation as the old variables. Moreover, we replace the expression for $P_{3\parallel}$ with the same expression divided by $m$. Similarly, the constants in expressions for cross-section, i.e. the designations $M$ and $\sqrt s$  are used for dimensionless proton mass and energy of colliding particles in c.m.s. (nondimensionalized with the pion mass $m$).

Next, substitute the following notations of integration variables into Eq.\ref{eq11} and designate the rapidities $y_1, y_2,\ldots, y_n$ as $X_1, X_2,\ldots, X_n$; $x$-components of transverse momenta of secondary particles $p_{1\perp x}$, $p_{2\perp x}$,$\ldots,p_{n\perp x}$ as $X_{n+1},X_{n+2},\ldots,X_{2n}$; $y$-components of transverse momenta of secondary $p_{1\perp y}$,  $p_{2\perp y}$, $\ldots,p_{n\perp y}$ as $X_{2n+1}$, $X_{2n+2}$, $\ldots, X_{3n}$. Finally, we define $X_{3n+1}$ as $P_{3\perp x}$ and $X_{3n+2}$ as $P_{3\perp y}$.

In the previous sections it has been shown that an integrand
$A(n,P_3,P_4,p_1,p_2,\ldots,p_n,P_1,P_2)$ in Eq.\ref{eq11}, expressed as a function of independent integration variables, has a maximum point in the domain of integration. In the vicinity of this
maximum point it can be represented in the form
\begin{eqnarray}
&& \mbox{\fontsize{13}{14}\selectfont $ A\left( {n,{P_3},{P_4},{p_1},{p_2},...,{p_n},{P_2},{P_1}} \right) = $ } \nonumber\\ 
&& \mbox{\fontsize{13}{14}\selectfont $ ={A^{\left( 0 \right),n}}\left( {\sqrt s } \right)$ } \nonumber\\ 
&& \mbox{\fontsize{9}{14}\selectfont $ \times \exp \left( { - \frac{1}{2}\sum\limits_{a = 1}^{3n + 2} {\sum\limits_{b = 1}^{3n + 2} {{D_{ab}}} \left( {{X_a} - X_a^{\left( 0 \right)}} \right)\left( {{X_b} - X_b^{\left( 0 \right)}} \right)} } \right)$ } \nonumber\\ 
\label{eq12} 
\end{eqnarray}
where ${A^{(0),\,n}}\left( {\sqrt s } \right)$ is the value of function (see Eq.4 in [\onlinecite{part1}]) at the point of constrained maximum; ${D_{ab}} =  - \frac{{{\partial ^2}}}{{\partial {X_a}\partial {X_b}}}\left( {\ln \left( A \right)} \right)$  $\textendash$ the derivatives are taken at the constrained maximum point of scattering amplitude; $X_a^{(0)}$ - value of variables, maximize the scattering amplitude. 
That is, the real and positive value $A$ defined by (see Eq.4 in [\onlinecite{part1}]) is represented as 
$A = \exp \left( {\ln \left( A \right)} \right)$, and exponential function is expanded into the Taylor series in the neighborhood of the maximum point with an accuracy up to the second-order summands.

An accuracy of approximation Eq.\ref{eq12} can be numerically verified in the following way. Function $A$, defined by (see Eq.4 [\onlinecite{part1})] can be written as 
\begin{eqnarray}
&& A = A\left( {n,{X_1},{X_2},...,{X_{3n + 2}}} \right) 
\label{eq13}
\end{eqnarray}

Let us introduce also the notation: 
\begin{eqnarray}
&& \mbox{\fontsize{12}{10}\selectfont $ {A^{\left( g \right)}}\left( {n,{X_1},{X_2},...,{X_{3n + 2}}} \right) = {A^{\left( 0 \right),n}}\left( {\sqrt s } \right)$ } \nonumber\\ 
&& \mbox{\fontsize{9}{8}\selectfont $ \times \exp \left( { - \frac{1}{2}\sum\limits_{a = 1}^{3n + 2} {\sum\limits_{b = 1}^{3n + 2} {{D_{ab}}} \left( {{X_a} - X_a^{\left( 0 \right)}} \right)\left( {{X_b} - X_b^{\left( 0 \right)}} \right)} } \right)$ } \nonumber\\ 
\label{eq14} 
\end{eqnarray}
\begin{figure*}
\begin{center}
\subfigure[]{
  \includegraphics[scale=0.42]{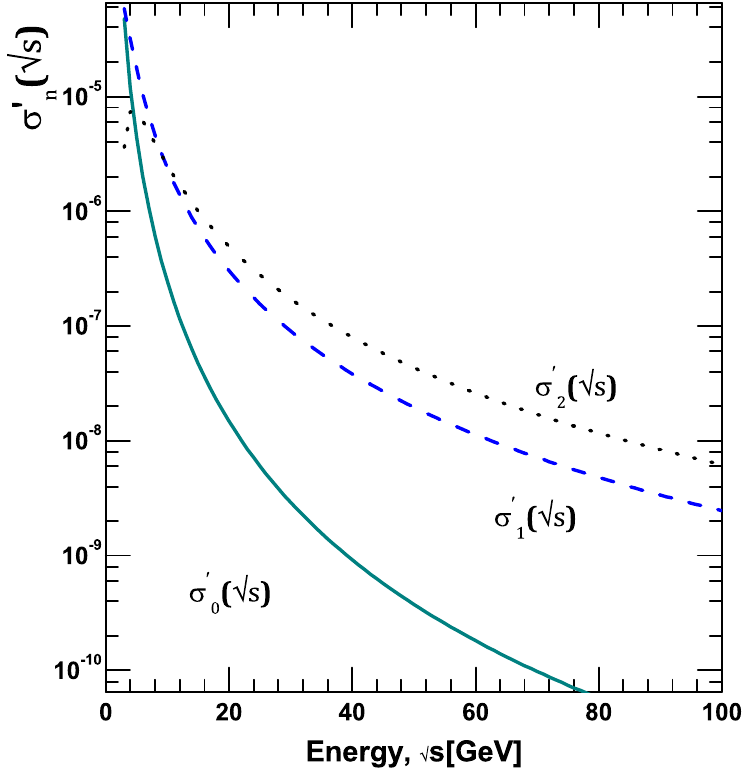}
  \label{fig:fig_part2_05a} 
}
\subfigure[]{
  \includegraphics[scale=0.42]{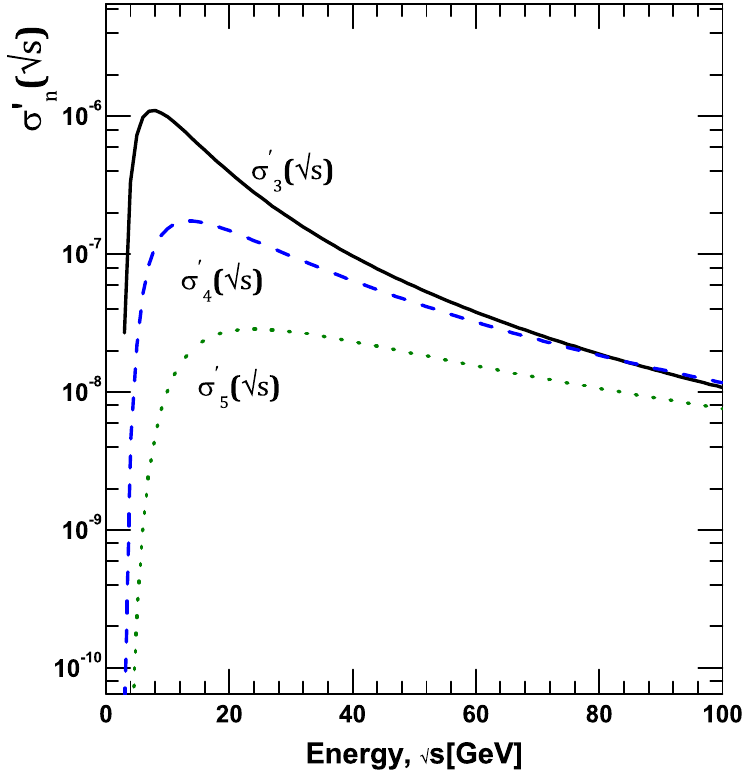} 
  \label{fig:fig_part2_05b} 
}
\subfigure[]{
  \includegraphics[scale=0.42]{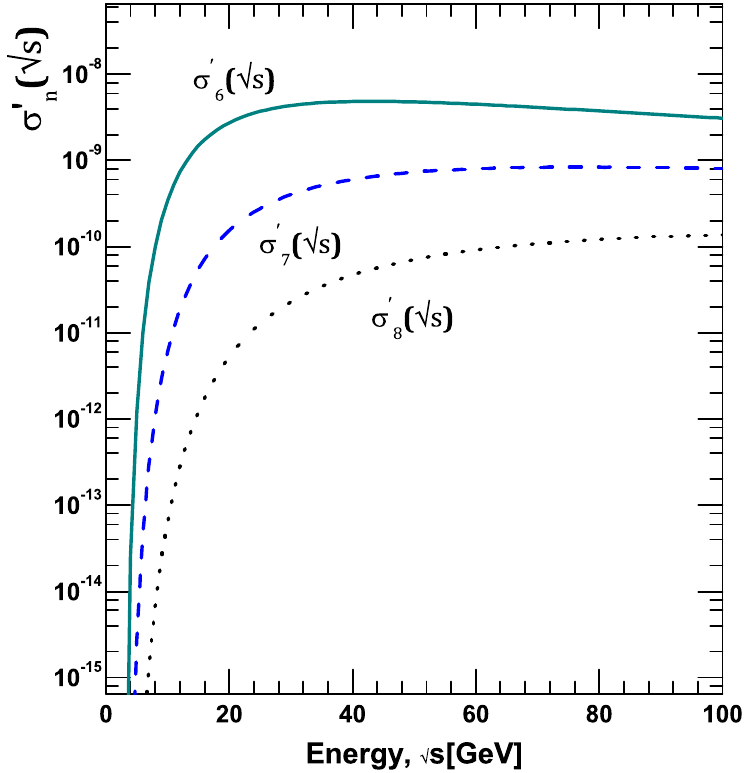} 
  \label{fig:fig_part2_05c} 
}
\end{center}
\caption{
The calculated values of ${\sigma '_n}\left( {\sqrt s } \right)$ for the energy range $\sqrt s  = 3 \div 95$ GeV.}
\label{fig:fig_part2_05}
\end{figure*}
Now let us examine functions:
\begin{eqnarray}
&& \mbox{\fontsize{13}{14}\selectfont $ F_{ik}^n\left( {a,b} \right) = $ } \nonumber\\ 
&& \mbox{\fontsize{9}{14}\selectfont $ =A\left( {n,X_1^{\left( 0 \right)},X_2^{\left( 0 \right)},...,X_i^{\left( 0 \right)} + a,...,X_k^{\left( 0 \right)} + b,...,X_{3n + 2}^{\left( 0 \right)}} \right)$ } \nonumber\\ 
\label{eq15a} 
\end{eqnarray}
\begin{eqnarray}
&& \mbox{\fontsize{13}{14}\selectfont $  F_{ik}^{\left( g \right),n}\left( {a,b} \right) =  $ } \nonumber\\ 
&& \mbox{\fontsize{8}{14}\selectfont $ ={A^{\left( g \right)}}\left( {n,X_1^{\left( 0 \right)},X_2^{\left( 0 \right)},...,X_i^{\left( 0 \right)} + a,...,X_k^{\left( 0 \right)} + b,...,X_{3n + 2}^{\left( 0 \right)}} \right)$ }\nonumber\\ 
\label{eq15b} 
\end{eqnarray}

Three-dimensional curves of these functions can be easily plotted in the vicinity of the maximum point (i.e., at the neighborhood of zero of variables $a$ and $b$). The typical examples of such curves are shown in Fig.\ref{fig:part2_fig02} and Fig.\ref{fig:part2_fig03}, where it is easy to see that the approximation Eq.\ref{eq14} works well in the wide energy range. Results similar to the plots in Figs.\ref{fig:part2_fig02}, \ref{fig:part2_fig03} were obtained at different values of $\sqrt s$, $i$, $k$ and $n$. As one can see from Fig.\ref{fig:part2_fig02} and Fig.\ref{fig:part2_fig03} the true amplitude and its Gaussian approximation in Eq.\ref{eq14} differ visibly only in the parameter region. This makes an insignificant contribution to the integral.

Now let us proceed with the identification $A(n,P_3,P_4,p_1,p_2,\ldots,p_n,P_1,P_2)$ in Eq.\ref{eq6} and define all possible arrangements of indices $1, 2,\ldots, n$ by $P^{(1)}, P^{(2)},\ldots, P^{(n!)}$. Let’s denote the function of variables $X_k$, where $k=1,2,\ldots,3n+2$, which corresponds to arrangement $P^{(l)}$, through
$A_{P^{(l)}}(n,X_1,\ldots,X_{3n+2})$. This function differs from the function in Eq.\ref{eq13} only by notaion, and therefore also has a constrained maximum point at the condition of the energy-momentum conservation. 
The value of this function at the constrained maximum point is equal to the value of function Eq.\ref{eq13}, i.e., and it is equal to $A^{(0),n}(\sqrt s)$ according to the replacement made above. Thus, if $X_1^{(0)}, X_2^{(0)},\ldots,X_n^{(0)}$ are the values of variables $X_1,X_2,\ldots,X_n$ at the maximum point, now same values $X_1^{(0)}, X_2^{(0)},\ldots,X_n^{(0)}$ will be the values of the variables $X_{j_1},X_{j_2},\ldots,X_{j_n}$ at the maximum point. 
Analogously $X_{n+1}^{(0)}, X_{n+2}^{(0)},\ldots,X_{2n}^{(0)}$ are the values of variables $X_{n+j_1},X_{n+j_2},\ldots,X_{n+j_n}$ at the maximum point, and $X_{2n+1}^{(0)}, X_{2n+2}^{(0)},\ldots,X_{3n}^{(0)}$ \textendash ~for $X_{2n+j_1},X_{2n+j_2},\ldots,X_{2n+j_n}$. For short, we denote the index of variable, into which the variable $a$ goes at given permutation, by $P^{(i)}(a)$ i.e., the variable $X_a$ gets replaced with $X_{P^{(l)}}(a)$.

Denoting the matrix of second derivatives of the function $A_{P^{(l)}}$ logarithm at the maximum point through $\hat{D}^{P^{(l)}}$, we get the following approximation for $A_{P^{(l)}}$:
\begin{eqnarray}
&& \mbox{\fontsize{10}{14}\selectfont $ {A_{{P^{\left( l \right)}}}}\left( {n,{X_1},{X_2},...,{X_{3n + 2}}} \right)= {A^{\left( 0 \right),n}}\left( {\sqrt s } \right)  \exp \left( - \frac{1}{2}K \right)$ }\nonumber\\
\label{eq17}
\end{eqnarray}
where,
\begin{eqnarray}
&& \mbox{\fontsize{12}{14}\selectfont $ K = \sum\limits_{ \scriptstyle a = 1 \hfill}^{3n + 2} \sum\limits_{ \scriptstyle b = 1 \hfill}^{3n + 2} {D_{{P^{\left( l \right)}}\left( a \right),{P^{\left( l \right)}}\left( b \right)}^{{P^{\left( l \right)}}}} $ }\nonumber\\
&& \mbox{\fontsize{12}{10}\selectfont $ \times \left( {{X_{{P^{\left( l \right)}}\left( a \right)}} - X_a^{\left( 0 \right)}} \right)\left( {{X_{{P^{\left( l \right)}}\left( b \right)}} - X_b^{\left( 0 \right)}} \right)$ }
\end{eqnarray}

Just as function $A$ depends upon variables $X_a$ and $X_b$, Eq.\ref{eq17} depends on variables $X_{P^{(l)}(a)}$ and $X_{P^{(l)}(b)}$. Therefore the second derivative is taken at the same values of arguments, and we have:
\begin{eqnarray}
&& D_{{P^{\left( l \right)}}\left( a \right),{P^{\left( l \right)}}\left( b \right)}^{{P^{\left( l \right)}}} = {D_{ab}} 
\label{eq19}
\end{eqnarray}
Using Eq.\ref{eq19} we rewrite Eq.\ref{eq17} in more convenient form. For this purpose introduce the matrices $\hat{P}^{(l)}$, $l=1,2,\ldots,n!$. Multiplying them by the column $\hat{X}$ of initial variables in Eq.\ref{eq14}, 
we get a column in which the variables are arranged such that in place of variable
$X_a$ became a variable $X_{P^{(l)}(a)}$. Next, taking into account Eq.\ref{eq19}, one can rewrite Eq.\ref{eq17} in matrix form:
\begin{eqnarray}
&& \mbox{\fontsize{10}{10}\selectfont $ {A_{{P^{\left( l \right)}}}}\left( {n,{X_1},{X_2},...,{X_{3n + 2}}} \right) = $}\nonumber\\ 
&& \mbox{\fontsize{10}{10}\selectfont $ ={A^{\left( 0 \right),n}}\left( {\sqrt s } \right) $} \nonumber\\ 
&& \mbox{\fontsize{8}{10}\selectfont $  \times \exp \left( { - \frac{1}{2}\left( {{{\hat X}^T}{{\left( {{{\hat P}^{\left( l \right)}}} \right)}^T}\hat D\,{{\hat P}^{\left( l \right)}}\hat X - 2{{\left( {{{\hat X}^{\left( 0 \right)}}} \right)}^T}\hat D\,{{\hat P}^{\left( l \right)}}\hat X} \right)} \right) $}\nonumber\\ 
&& \mbox{\fontsize{9}{10}\selectfont $  \times \exp \left( { - \frac{1}{2}\left( {{{\left( {{{\hat X}^{\left( 0 \right)}}} \right)}^T}\hat D\,{{\hat X}^{\left( 0 \right)}}} \right)} \right) $}
\end{eqnarray}
where $\hat{X}^{(0)}$ is a column vector whose elements are the numbers $X_a^{(0)}, a=1,2,\ldots,3n+2$, in the initial arrangement. 
Eq.\ref{eq6} can be rewritten in the form:
\begin{eqnarray}
&& \mbox{\fontsize{12}{14}\selectfont $ \Phi \left( {n,{P_3},{P_4},{p_1},{p_2}, \cdots ,{p_n},{P_2},{P_1}} \right) $ } \nonumber\\ 
&& \mbox{\fontsize{12}{14}\selectfont $  = \Phi \left( {n,{X_1},{X_2}, \cdots ,{X_{3n + 2}}} \right) $ } \nonumber\\ 
&& \mbox{\fontsize{12}{14}\selectfont $  = {\left( {{A^{\left( 0 \right),n}}\left( {\sqrt s } \right)} \right)^2}\exp \left( { - {{\left( {{{\hat X}^{\left( 0 \right)}}} \right)}^T}\hat D\,{{\hat X}^{\left( 0 \right)}}} \right) $ } \nonumber\\ 
&& \mbox{\fontsize{12}{14}\selectfont $  \times \sum\limits_{l = 1}^{n!} {\exp } \left( { - \frac{1}{2}{{\hat X}^T}{{\hat D}^{\left( l \right)}}\hat X + {{\left( {{{\hat X}^{\left( 0 \right)}}} \right)}^T}{{\hat V}^{\left( l \right)}}\hat X} \right) $ } \nonumber\\ 
\end{eqnarray}
where 
\begin{subequations}
\begin{eqnarray}
&& {\hat D^{\left( l \right)}} = \hat D + {\left( {{{\hat P}^{\left( l \right)}}} \right)^T}\hat D{\hat P^{\left( l \right)}} \\
&& {\hat V^{\left( l \right)}} = \hat V + \hat V{\hat P^{\left( l \right)}}
\end{eqnarray}
\end{subequations}

If now we turn to Eq.\ref{eq11}, we can see that all of the coefficients (except $\Phi'$) under the integration sign don’t change their values under permutation of the arguments. We replace these expressions by their values at the maximum point and take them out from integral. From this, we introduce the following notation:
\begin{eqnarray}
&& {J^{\left( 0 \right),n}}\left( {\sqrt s } \right) = {\left. {\frac{{\partial {P_{3\parallel }^{(0)}}}}{{\partial {E_p}}}} \right|_{{E_p} = \sqrt s  - \sum\limits_{k = 1}^n {ch\left( {y_k^{\left( 0 \right)}} \right)} }}
\end{eqnarray}
where $P_{3\parallel}^{(0)}$ is given in (see Eq.8 in [\onlinecite{part1}]) which corresponds to particles momenta maximizing the scattering amplitude. That is, the value at $X_a^{(0)}$ nondimensionalized with the mass $m$.

The expression for cross-section in this case can be written in the form:
\begin{eqnarray}
&& \mbox{\fontsize{12}{10}\selectfont $ {\sigma _n} = \frac{{{{\left( {2\pi } \right)}^2}}}{{16{m^2}\sqrt {s/4 - {M^2}} \sqrt s \sqrt {{M^2} + {{\left( {P_{3\parallel }^{\left( 0 \right)}} \right)}^2}} }} $ } \nonumber\\ 
&& \mbox{\fontsize{11}{10}\selectfont $  \times \frac{1}{{\sqrt {{M^2} + {{\left( {\sum\limits_{k = 1}^n {{\mathop{\rm sh}\nolimits} \left( {y_k^{\left( 0 \right)}} \right)}  + P_{3\parallel }^{\left( 0 \right)}} \right)}^2}} }} {\left( {\frac{g}{m}} \right)^4}{\left( {\frac{1}{{2{{\left( {2\pi } \right)}^3}}}{{\left( {\frac{\lambda }{m}} \right)}^2}} \right)^n} $ } \nonumber\\ 
&& \mbox{\fontsize{9}{10}\selectfont $ \times {\left( {{A^{\left( 0 \right),n}}\left( {\sqrt s } \right)} \right)^2}{J^{\left( 0 \right),n}}\left( {\sqrt s } \right)\exp \left( { - {{\left( {{{\hat X}^{\left( 0 \right)}}} \right)}^T}\hat D\,{{\hat X}^{\left( 0 \right)}}} \right)$ } \nonumber\\ 
&& \mbox{\fontsize{9}{10}\selectfont $ \times \sum\limits_{l = 1}^{n!} {\int {\prod\limits_{k = 1}^{3n + 2} {d{X_a}} } } \exp \left( { - \frac{1}{2}{{\hat X}^T}{{\hat D}^{\left( l \right)}}\hat X + {{\left( {{{\hat X}^{\left( 0 \right)}}} \right)}^T}{{\hat V}^{\left( l \right)}}\hat X} \right)$ }\nonumber\\ 
\label{eq26}
\end{eqnarray}

Since $\sum\limits_{k=1}^n sh(y_k^{(0)})+P_{3\parallel}^{(0)}$ in Eq.\ref{eq26} is the negative value of the longitudinal component of momentum $P_{4\parallel}^{(0)}$ taken at the maximum point, it can be replaced by $P_{3\parallel}^{(0)}$ due to the symmetry properties that have already been discussed.

Multi-dimensional integrals under the summation sign can be calculated by diagonalizing the quadratic form in the
exponent of each of them. Such diagonalization can be numerically realized, for instance, by the Lagrange method.
Calculating the large number of terms in Eq.\ref{eq26} is a substantial computational problem, which we overcome only for $n\leq 8$.

To represent the results of those numerical computations, it is useful to decompose Eq.\ref{eq26} in the following way:
\begin{eqnarray}
&& f_p^{\left( n \right)}\left( {\sqrt s } \right) = \exp \left( { - {{\left( {{{\hat X}^{\left( 0 \right)}}} \right)}^T}\hat D{{\hat X}^{\left( 0 \right)}}} \right) \nonumber\\ 
&& \mbox{\fontsize{7}{10}\selectfont $ \times \sum\limits_{l = 1}^{n!} {\int {\prod\limits_{k = 1}^{3n + 2} {d{X_a}} } \exp \left( { - \frac{1}{2}{{\hat X}^T}{{\hat D}^{\left( l \right)}}\hat X + {{\left( {{{\hat X}^{\left( 0 \right)}}} \right)}^T}{{\hat V}^{\left( l \right)}}\hat X} \right)}$ } 
\label{eq27} \\
&& \mbox{\fontsize{12}{10}\selectfont $ {\sigma '_n}\left( {\sqrt s } \right) = \frac{{{{\left( {{A^{\left( 0 \right),n}}\left( {\sqrt s } \right)} \right)}^2}{J^{\left( 0 \right),n}}\left( {\sqrt s } \right)f_p^{\left( n \right)}\left( {\sqrt s } \right)}}{{\sqrt {s/4 - {M^2}} \sqrt s \left( {{M^2} + {{\left( {P_{3\left\| {} \right.}^{\left( 0 \right)}} \right)}^2}} \right)}}$ } 
\label{eq27a} \\
&& L = \frac{1}{{2{{\left( {2\pi } \right)}^3}}}{\left( {\frac{\lambda }{m}} \right)^2} 
\label{eq29} 
\end{eqnarray}

From now on we employ the ``prime” sign in our notations to indicate that we are using a dimensionless quantity, which characterized the dependence of the cross-sections on energy, but not their absolute values.

Eq.\ref{eq27a} differs from the inelastic scattering cross-section $\sigma' _n(\sqrt s)$ only by the absence of factor $\frac{\left(2\pi \right)^2}{16m^2}\left( \frac{g}{m} \right)^4\left( \frac{1}{2(2\pi)^3} \left( \frac{\lambda}{m} \right)^2 \right)^n$, which is energy independent. An investigation of Eq.\ref{eq27a} allows us to trace the dep pendence of inelastic scattering cross-section on energy $\sqrt s$ (Fig.\ref{fig:fig_part2_04} and Fig.\ref{fig:fig_part2_05}).

From Fig.\ref{fig:fig_part2_04} it is obvious that derivatives of cross-sections with respect to energies along the real axis are equal to zero at points corresponding to the threshold energy of $n$ particle production.
In other words, although the threshold values of energy are the branch points of crosssections, the cross-sections indeed have continuous first derivatives along the real axis at these branch points.
This can be illustrated in the following way.
In the examined approximation of equal denominators, for the even number of particles value of square of scattering amplitude at the maximum point can be written like:
\begin{eqnarray}
&& \mbox{\fontsize{13}{10}\selectfont ${\left( {{A^{\left( 0 \right),n}}} \right)^2} = {\left( {1 + \frac{1}{{{{{\mathop{\rm sh}\nolimits} }^2}\left( {{y_{\frac{n}{2}}}} \right)}}} \right)^{ - 2\left( {n + 1} \right)}}$}
 \label{eq2_19} 
\end{eqnarray}
where ${y_{\frac{n}{2}}}$ defined by (see \onlinecite{part1}):
\begin{eqnarray}
&& \mbox{\fontsize{13}{10}\selectfont ${y_{\frac{n}{2}}} = \frac{1}{{n + 1}}{\mathop{\rm acosh}\nolimits} \left( {\frac{{\sqrt s  - n}}{{2M}}} \right)$}
 \label{eq2_19a} 
\end{eqnarray}
Derivative from Eq.\ref{eq2_19a} along the real axis at the threshold branching-point is infinite. However, cause at this point value of ${y_{\frac{n}{2}}}$ is zero, than from Eq.\ref{eq2_19} it is obvious that derivative of  ${{A^{\left( 0 \right),n}}}$ will be converge to threshold along the real axis tends to zero.

\begin{figure}
\begin{center}
\includegraphics[scale=0.33]{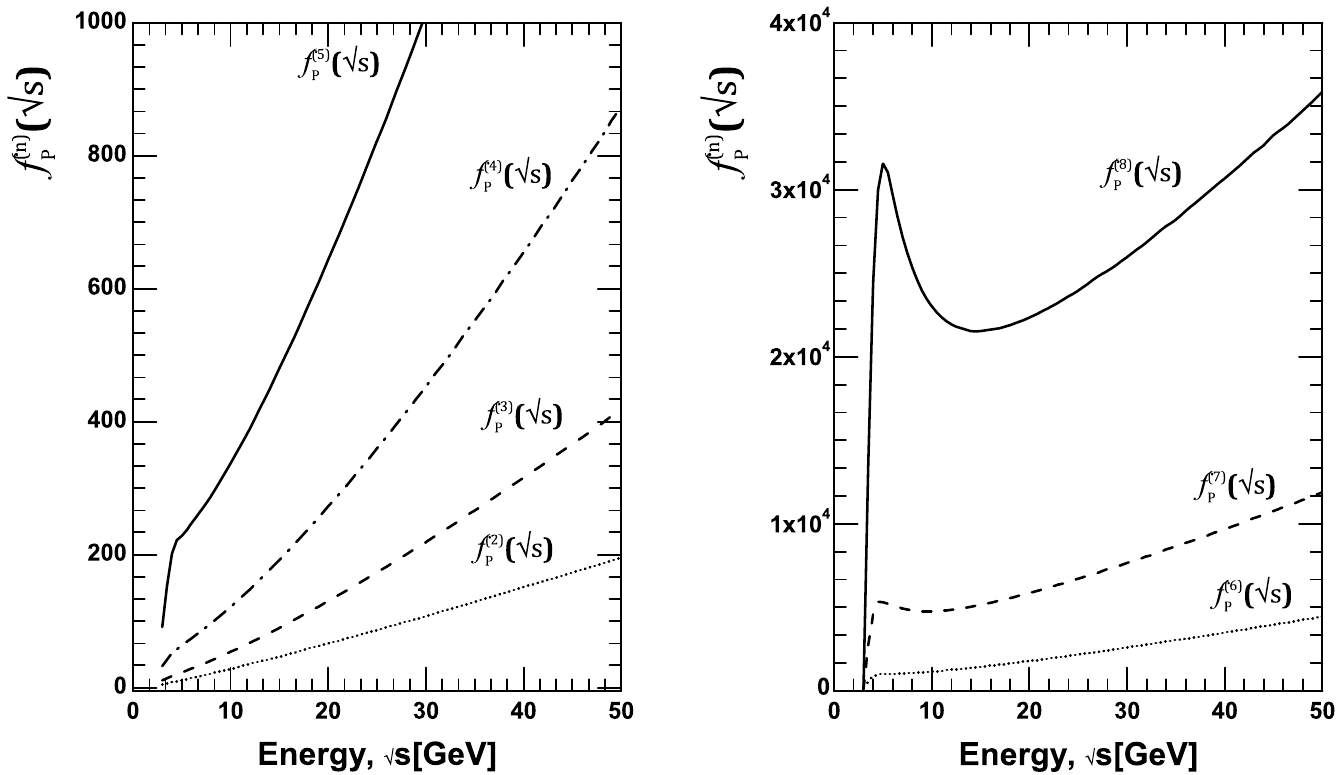}
\end{center}
\caption{
The calculated values of $f_P^{\left( n \right)}\left( {\sqrt s } \right)$ determined by Eq.\ref{eq27} for the energy range $\sqrt s  = 3 \div 95$ GeV. }
\label{fig:fig_part2_06}
\end{figure}

As it follows from Figs.\ref{fig:fig_part2_04}-\ref{fig:fig_part2_05}, $\sigma '_8(\sqrt s)$ monotone increases in the all considered energy range. At the same time from Fig.\ref{fig:fig_part2_06} one can see that $f_P^{(8)}(\sqrt s)$ has drop-down sections. Moreover, even on those sections, where $f_P^{(n)}(\sqrt s), n=2\div5$ increase, corresponding $\sigma '_n(\sqrt s)$ decrease. It makes possible to conclude, that amplitude growth at maximum point (which is the consequence of virtuality reduction) is generally responsible for the growth of inelastic scattering cross-section.

\begin{figure}
\begin{center}
\includegraphics[scale=0.42]{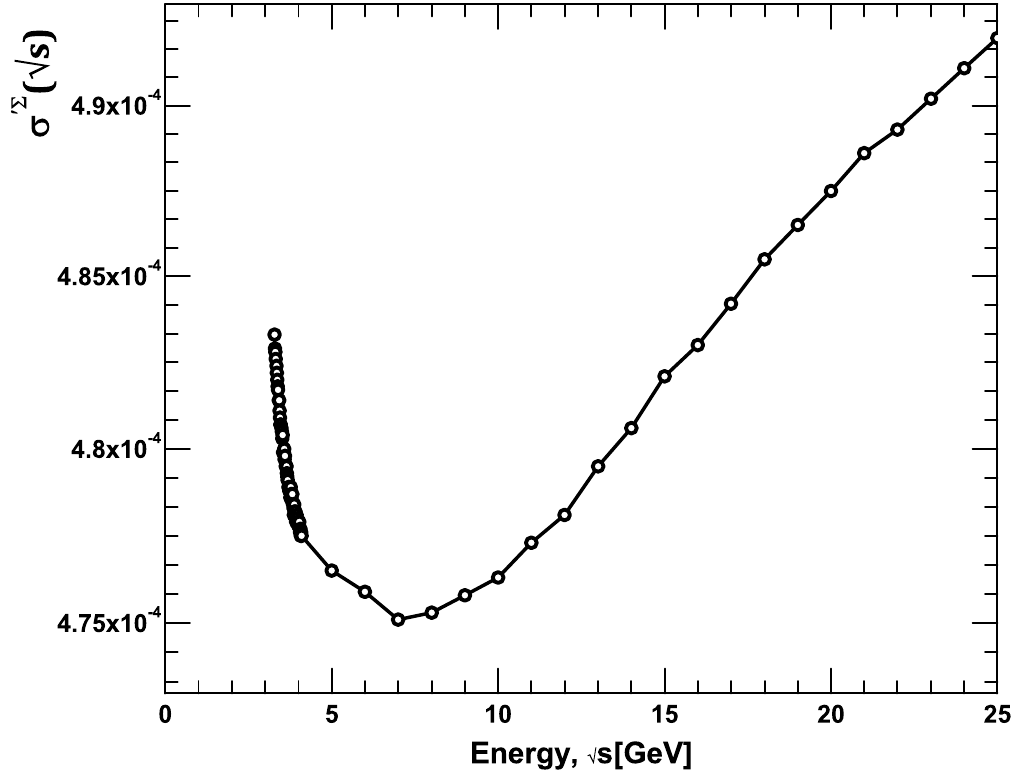}
\end{center}
\caption{
Calculated values of ${\sigma '^{\sum {} }}\left( {\sqrt s } \right)$ at $L = 5.57$, in the energy range $\sqrt s  = 5 \div 25$ GeV. }
\label{fig:fig_part2_07}
\end{figure}

\begin{figure}
\begin{center}
\subfigure[]{
  \includegraphics[scale=0.42]{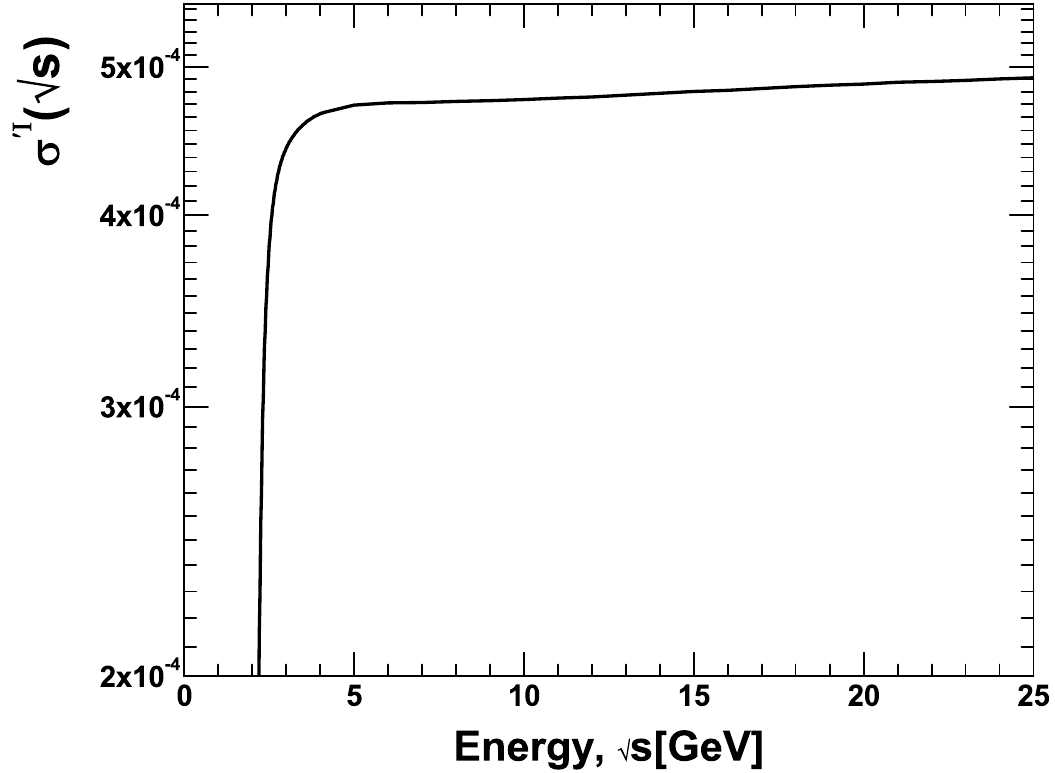}
  \label{fig:fig_part2_08a} 
}
\subfigure[]{
  \includegraphics[scale=0.42]{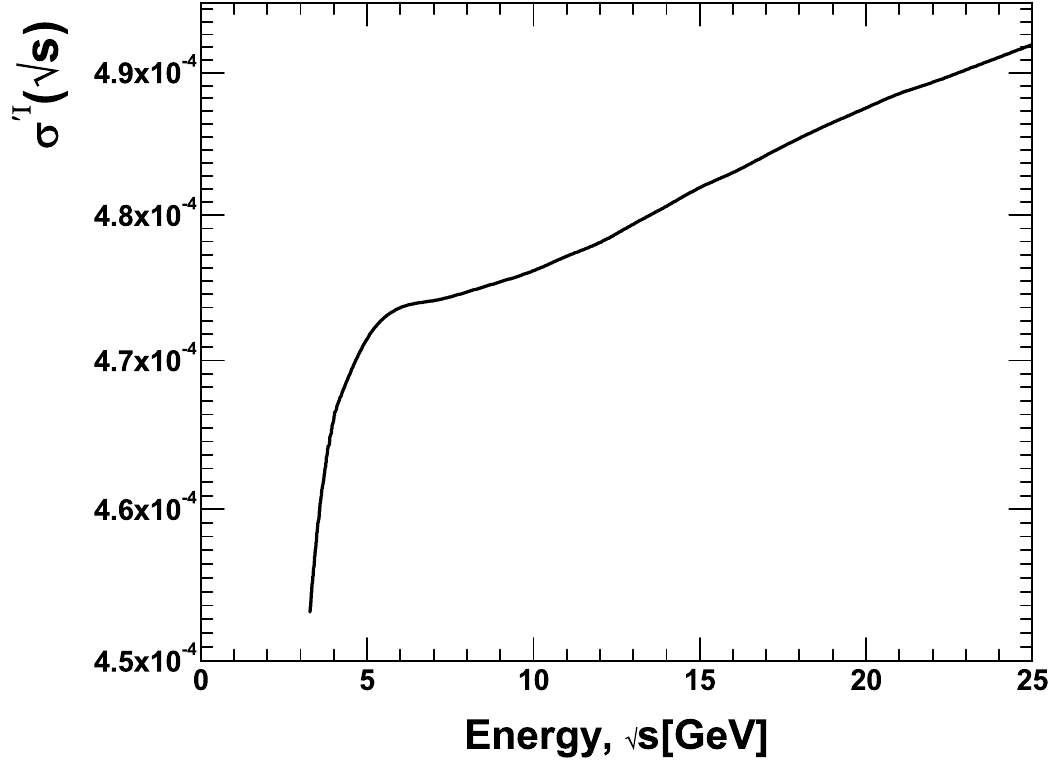} 
  \label{fig:fig_part2_08b} 
}
\end{center}
\caption{
Calculated values of ${\sigma '^I}\left( {\sqrt s } \right)$ at $L = 5.57$, in the energy range: a) $\sqrt s  = 1.89 \div 25$ GeV 
(starting from the threshold of inelastic scattering); 
b) $\sqrt s  = 3 \div 25$ GeV (where the increase of total cross-section with energy growth is clearly visible). }
\label{fig:fig_part2_08}
\end{figure}

As it evident from Fig.\ref{fig:fig_part2_04} and Fig.\ref{fig:fig_part2_05} for energy values ​​in Eq.\ref{eq27a} has a positive energy derivative, and other enrgy values have a negative derivative.
Thus the question arises: if we take their sums 
\begin{eqnarray}
&& {\sigma '^\Sigma }\left( {\sqrt s } \right) = \sum\limits_{n = 0}^8 {{L^n}{{\sigma '}_n}\left( {\sqrt s } \right)} 
\label{eq29a}
\end{eqnarray} 
and 
\begin{eqnarray}
&& {\sigma '^I}\left( {\sqrt s } \right) = \sum\limits_{n = 1}^8 {{L^n}{{\sigma '}_n}\left( {\sqrt s } \right)} 
\label{eq29b}
\end{eqnarray}
where $L$ is defined by Eq.\ref{eq29}, then is it possible to choose the ``coupling constant" $L$ such that the value of Eq.\ref{eq29a} has a characteristic minimum similar to the one observed experimentally for total proton-proton scattering cross-section?
The answer is yes (see Fig.\ref{fig:fig_part2_07}). We find that curves agree qualitatively at the close values of $L$. The energy range shown in Fig.\ref{fig:fig_part2_07} takes into account all the inelastic contributions. We find indeed very interesting result, that curves presented on Fig.\ref{fig:fig_part2_07} and on Fig.\ref{fig:fig_part2_08}, where calculated values of Eqs.\ref{eq29a}, \ref{eq29b} are given at $L=5.57$, qualitatively agree with experimental data [\onlinecite{PDG_2010_JofPhysG, ATLASCollaboration_2011eu}]. Quantitative agreement was not achieved! Here, we would like to emphasize this fact, due to avoid any speculations.
Furthermore, let us point to the fact that in Fig.\ref{fig:fig_part2_07}-\ref{fig:fig_part2_08}
the minimum at higher energies $\sqrt s $ than in the experiment. We believe that the accounting contributions with higher number of secondary particles $n$ to ${\sigma '_n}\left( {\sqrt s } \right)$ and the corresponding change of constant $L$ will ``move'' a maximum to a required area.

Moreover, in this paper we have examined the simplest diagrams of $\phi^3$ theory and we intend to compare the qualitative form of these cross-sections with experimental data, but do not claim quantitative agreement. It is possible to hope that the application of similar computation method to more complicated diagrams in more realistic models will lead to correct outcome.

As known, within the framework of Reggeon theory the drop-down part of total cross-section is described by the reggeons exchanges with interception less than unity [\onlinecite{Donnachie1992227, Kaidalov:2003}]. The cuts concerned with multi-Reggeon exchanges with participation of reggeons with intercept greater than unity are responsible for the cross-section growth after the reaching the minimum [\onlinecite{Collins:111502}].

As will be shown further, the account for $\sigma '_n(\sqrt s)$ at $n>8$ will not change the behavior of function $\sigma '^{\Sigma}(\sqrt s)$ Eq.\ref{eq29a}.
Within the framework of the given model, the summation of multi-peripheral diagrams within the calculation of the imaginary part of elastic scattering amplitude will not result in power dependence on energy, since this dependence is monotonic. This means that the corresponding partial amplitude does not have a pole! This obviously differs from the results of standard approach in calculations of multi-peripheral model and from the results of Reggeon theory (see f.ex. [\onlinecite{springerlink:10.1007/BF02781901}]). 
 
\begin{figure}
\begin{center}
\includegraphics[scale=0.46]{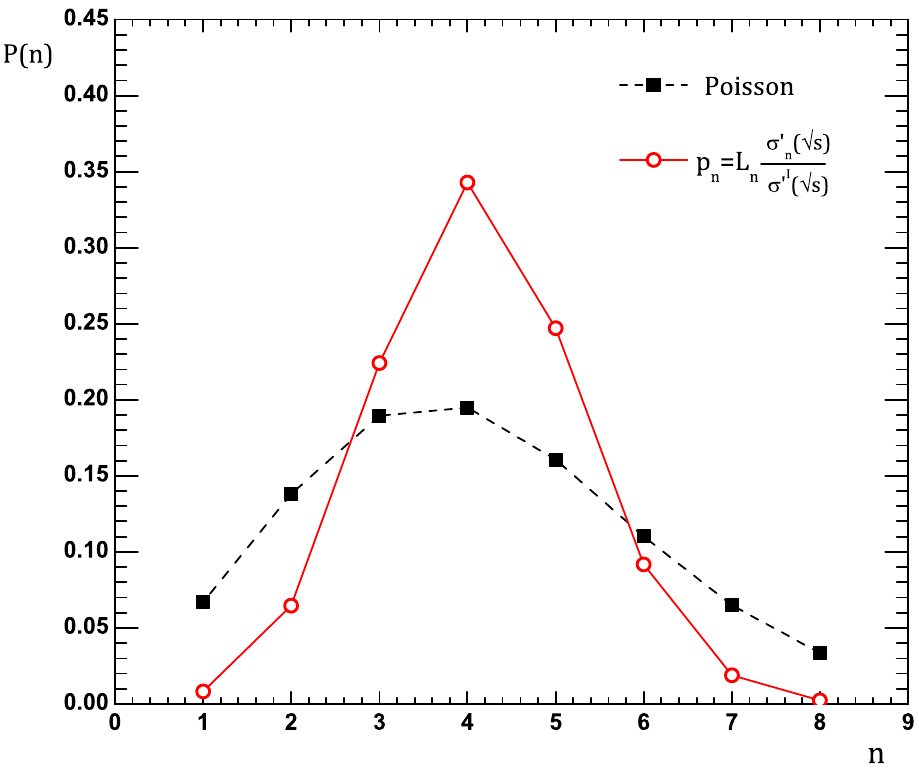}
\end{center}
\caption{
Distribution obtained from Eq.\ref{eq31} (red line) in comparison with Poisson distribution (dotted line) at $\sqrt s  = 15$ GeV. }
\label{fig:fig_part2_09}
\end{figure}

Another argument in favor of this hypothesis are the results of the ``multiplicity distribution" shown in Fig.\ref{fig:fig_part2_09}, where axis of ordinates designates the number of particles $n$ and abscissa axis designates the value of:
\begin{eqnarray} 
&& {p_n} = \frac{{{L^n}{{\sigma '}_n}\left( {\sqrt s } \right)}}{{{{\sigma '}^I}\left( {\sqrt s } \right)}} 
\label{eq31}
\end{eqnarray}

The Poisson distribution for the same average like for distribution Eq.\ref{eq31} is given for comparison. The energy $\sqrt s  = 15$ GeV is chosen for example, because at higher energies all distributions is no longer fit in the range from 0 to 8 particles. As is obvious from Fig.\ref{fig:fig_part2_09}, the distribution Eq.\ref{eq31} significantly differs from the Poisson distribution, which, as it is known, lead to power-law behavior of the imaginary part of inelastic scattering amplitude and, consequently, to the pole singularity of partial amplitude [\onlinecite{Nikitin:113716, Collins:111502}].

The described differences from a Regge theory are caused, apparently, by different physical mechanisms determining the inelastic scattering cross-section growth. In our model, a reduction of virtualities at the point of constrained maximum of inelastic scattering amplitude play a role of such mechanism. Consideration of similar diagrams in [\onlinecite{springerlink:10.1007/BF02781901}] lead to
\begin{eqnarray}  
&& {\sigma '_n} \approx \frac{1}{{n!}}\frac{{{{\ln }^n}\left( s \right)}}{{{s^2}}} 
\label{eq32}
\end{eqnarray}

At the same time a similar result is obtained in [\onlinecite{Byckling:100542}] by the calculating of phase space with ``cutting'' of transversal momenta, i.e. authors ignore the dependence of inelastic scattering amplitude on rapidity, and its role is reduced only to the cutting of integration over transversal momenta. Similar results are obtained in [\onlinecite{Nikitin:113716, Collins:111502}], where examined diagrams of same type, but with the exchange of reggeons instead of virtual scalar particles was considered. In [\onlinecite{Nikitin:113716, Collins:111502}] as a result of approximation authors totally ignored the dependence of expression under the integral sign for cross-section on particle rapidity in the final state, thus obtained results include the dependence on energy $\sqrt s$ only through the rapidity phase space. At the same time, as it evident from previous argumentations, the dependence of scattering amplitude on longitudinal momenta or rapidity is essential, because it is responsible for the certain mechanism of inelastic cross-sections growth and their sum.

Moreover, Eq.\ref{eq32} has positive derivative with respect to energy at sufficiently great $n$ in sufficiently wide energy range. At the same time, sum of such expressions in [\onlinecite{springerlink:10.1007/BF02781901}] results in the cross-section, which decreases monotonically with energy growth. The reason for this may be apparent from Eq.\ref{eq32}  factorial suppression of contributions with large $n$, which provide the positive contributions to derivative with respect to energy. 

\begin{figure}
\begin{center}
\includegraphics[scale=0.5]{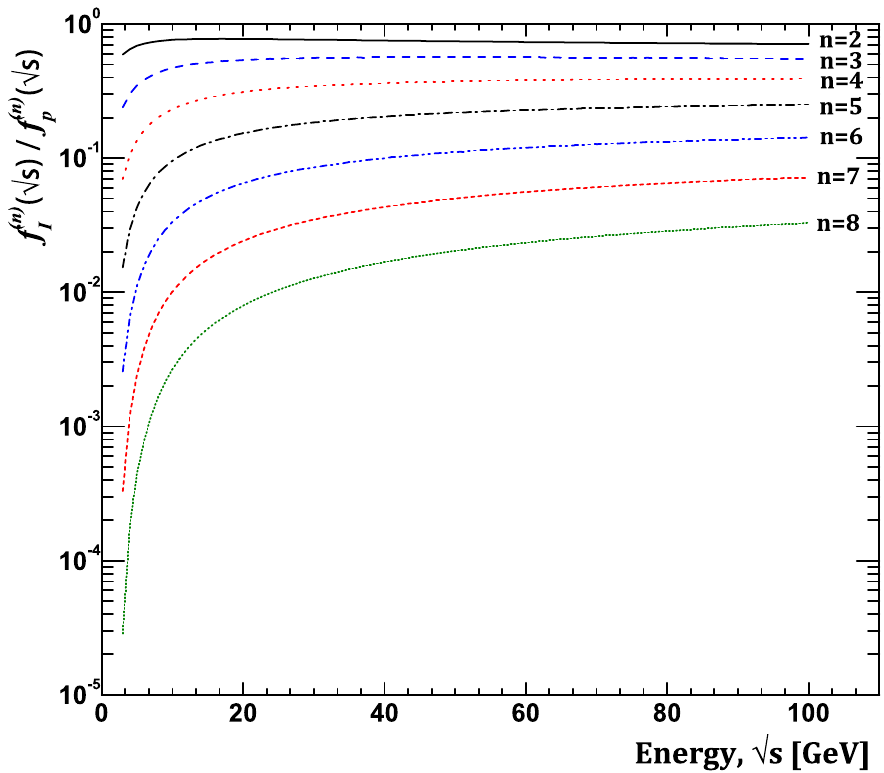}
\end{center}
\caption{
The ratio of the contribution from diagrams with an initial arrangement of the momenta $f_I^{(n)}\left( {\sqrt s } \right)$ to the sum of diagrams corresponding to all possible momenta arrangements $f_p^{(n)}\left( {\sqrt s } \right)$ at different $n$. }
  \label{fig:fig_part2_10}
\end{figure}

In the presented model such suppression disappears at transition from Eq.\ref{eq2} to Eq.\ref{eq5} due to taking into account diagrams, with different order of attachment of external lines to the ``comb''. The fact that the inclusion of such diagrams is essential as it seen from Fig.\ref{fig:fig_part2_10}, where the ratio of contribution $f_I^{\left( n \right)}\left( {\sqrt s } \right)$ from a diagram with the initial arrangement of momenta (see Fig.2 in [\onlinecite{part1}]) corresponding to the first summand in a sum Eq.\ref{eq27} to all sum $f_P^{\left( n \right)}\left( {\sqrt s } \right)$ is given. 

As seen from Fig.\ref{fig:fig_part2_10} contribution from a diagram with the initial arrangement of external lines in the wide energy range is small fraction of the total sum Eq.\ref{eq27}, which was natural to expect since sum Eq.\ref{eq27} has enormous number of positive summands. For the same reason, as was shown on Fig.\ref{fig:fig_part2_10}, the quota of contribution from a diagram with the initial arrangement of particles decreases sharply with increasing number of particles $n$ in a ``comb''.

At the same time, as it follows from Eq.\ref{eq27a}, the growth of scattering amplitude at the maximum point related with the mechanism of reduction of virtualities can cause the growth of inelastic scattering cross-sections ${\sigma' _n}\left( {\sqrt s } \right)$ and, consequently, the growth of total cross-section. As an argument we can show results of numerical calculation of the function Eq.\ref{fun}, which are listed in Table.\ref{part2_Table1}. 
\begin{eqnarray}
&& \mbox{\fontsize{13}{14}\selectfont $ {Q_n}\left( {\sqrt s } \right) = \frac{{{{\left( {{A^{\left( 0 \right),n}}\left( {\sqrt s } \right)} \right)}^2}{J^{\left( 0 \right),n}}\left( {\sqrt s } \right)}}{{\sqrt {s/4 - {M^2}} \sqrt s \left( {{M^2} + {{\left( {P_{3\parallel }^{\left( 0 \right)}} \right)}^2}} \right)}}$ } 
\label{fun}
\end{eqnarray}

This function is the ratio of increasing amplitude at the maximum point to the multipliers, which ``working'' on lowering of the total cross-sections with energy growth.

\begin{table}[h]
\begin{ruledtabular}
\begin{tabular}{cccccccc}
$\sqrt s $, GeV& $\ln ({Q_{10}}(s))$ & $\ln ({Q_{20}}(s))$ \\
\hline
5&	-68.867&	-202.469\\
15&	-48.936&	-133.814\\
25&	-44.874&	-120.196\\
35&	-43.036&	-113.138\\
45&	-41.993&	-108.585\\
55&	-41.328&	-105.315\\
65&	-40.874&	-102.81\\
100&	-40.065&	-97.131\\
200&	-39.622&	-89.901\\
300&	-39.74&	-86.537\\
500&	-40.191&	-83.029\\
900&	-41.043&	-79.833\\
1800&	-42.399&	-77.008\\
5000&	-44.905&	-74.318\\
14000&	-47.862&	-72.979\\
\end{tabular}
\end{ruledtabular}
\caption{ 
Energy dependence of the function Eq.\ref{fun} at $n=10$ and $n=20$.}
\label{part2_Table1}
\end{table}
Submitted data shows that the mechanism of virtuality reduction is ``stronger'' than multipliers, which ``working'' on lowering of the total cross-sections with energy growth.

From Eq.\ref{eq2_19} follows that with increasing of $n$ amplitude at the maximum point will increase sharply with energy growth. Thus, we can expect that factor $f_P^{\left( n \right)}\left( {\sqrt s } \right)$, which besides of ${Q_n}\left( {\sqrt s } \right)$ also inters into the expression of cross-section will decrease, but quite slowly.
As it obvious from Eq.\ref{eq6}, the possible decrease of $f_P^{\left( n \right)}\left( {\sqrt s } \right)$ is caused due to the fact that $f_P^{\left( n \right)}\left( {\sqrt s } \right)$ include the product of terms, corresponding to diagrams in which external lines with the same momenta can be attached to the different vertices of the diagram. As result, the momentum of such line can not have a value that simultaneously set maximum for both vertices. Moreover, with energy growth distance between rapidities corresponding to particles, which providing maximum at the different vertices of the diagram, increase.  This can lead to decreasing of value $f_P^{\left( n \right)}\left( {\sqrt s } \right)$ with energy.
However, as it obvious from relations (see Eq.75 and Eq.81 [\onlinecite{part1}]), we write them here:
\begin{eqnarray}
&& \mbox{\fontsize{13}{10}\selectfont ${y_{\frac{n}{2}}} = \frac{1}{{n + 1}}{\mathop{\rm acosh}\nolimits} \left( {\frac{{\sqrt s  - n}}{{2M}}} \right)$}
 \label{eq_part1_eq4_33} 
\end{eqnarray}
\begin{eqnarray}
&& \mbox{\fontsize{13}{10}\selectfont ${y_{\frac{n-1}{2}}} = \frac{2}{{n + 1}}{\mathop{\rm acosh}\nolimits} \left( {\frac{{\sqrt s  - n}}{{2M}}} \right)$}
 \label{eq_part1_eq4_39} 
\end{eqnarray}
the difference of these rapidities decreases with increase of particle`s number on the diagram. 
Therefore, it is hoped that decrease of $f_P^{\left( n \right)}\left( {\sqrt s } \right)$, even if it will take place, will be not too sharp and cross-sections for high multiplicities of particles will also grow at least in the certain energy range. This will lead to the amplification of contributions with positive derivative with respect to energy into the total scattering cross-section.

As it follows from Eq.\ref{eq2_19}, that at sufficiently high energies the amplitude at the maximum point tends to a constant value and mechanism of the reduction of virtualities become exhausted. This, however, can be avoided if we consider model in which the virtual particles on the diagram of the ``comb'' type are field quanta with zero mass. Then amplitude at the maximum point will tends to infinity at the infinite increase of energy. All computation in this case can be done similarly to what was described above. In this case, when calculating the first eight inelastic contributions in the wide range of energies does not give us contributions with negative derivative with respect to energy. Therefore we inclined to believe that such model can describe total cross-section growth to arbitrary large energies.

\section{Conclusions}
\label{SECTION_CONCLUSIONS}
From demonstrated results it can be conclude that replacing of the ``true'' scattering amplitude associated to the multi-peripheral processes within the framework of perturbation theory by its Gaussian approximation is an acceptable approximation. The main conclusion is, that the mechanism of virtuality reduction (considered in [\onlinecite{part1}]) may play a major role in ensuring the experimentally observed increase of the total cross-section [\onlinecite{PDG_2010_JofPhysG, ATLASCollaboration_2011eu}], at least in some range of energies. This growth was obtained with allowance for ${\sigma _n}$ at $n\leq 8$. However, as it follows from ${\sigma '_n}\left( {\sqrt s } \right)$ dependences, the maximum point of cross-section is shifted toward to higher energies with increase of $n$.
We can therefore expect that in the consider energy range accounting of ${\sigma '_n}\left( {\sqrt s } \right)$ will add summands with positive derivative with respect to energy to expression for the total scattering cross-section, which leads to the fact that at least in the considered energy range obtained growth will only intensify. 

Discussed above differences from the Reggeon theory suggest that our model is not a model of reggeon with intercept high than unity and increase of the cross section is occurred in different way. This is also evident from the fact that in the model with a nonzero mass of virtual particles cross-section $\sigma '_n$ at $\sqrt s \rightarrow \infty$ tends to zero. This is a consequence of the fact that the absolute value of virtualities can not decrease indefinitely, because it is bounded below by zero. Therefore, for sufficiently low coupling, when the arbitrarily high multiplicities $n$ do not contribute to the total cross section, at sufficiently high energies the total cross section should begin to decrease.

An additional conclusion is the necessity of accounting the sum of all diagrams with all the permutations of external lines for the scattering amplitude. Although with energy growth the fraction of contribution to the cross section of the diagram with an initial arrangement of the lines of the final particles increases and with $\sqrt s \rightarrow \infty$ will tends to unity. In a wide range of energies, this fraction is small and decreases with multiplicity increase $n$, which can be easily understood on the basis of the positivity of the amplitudes in the multi-peripheral model.

Note, that the application of Laplace method is not limited by simplest diagrams. Therefore, our goal is further consideration of the more realistic models using  same method, especially in terms of the law of conservation of electric charge.

\vspace{2cm} 
\normalsize {\textbf{REFERENCES}}
\bibliography{References_JMPh}

\end{document}